\documentclass{article}

\usepackage{PRIMEarxiv}

\usepackage{float}          
\usepackage[utf8]{inputenc} 
\usepackage[T1]{fontenc}    
\usepackage{hyperref}       
\usepackage{url}            
\usepackage{booktabs}       
\usepackage{amsfonts}       
\usepackage{nicefrac}       
\usepackage{microtype}      
\usepackage{graphicx}       
\graphicspath{{media/}}     
\usepackage{array}
\usepackage{amsmath}
\usepackage{enumitem}

\title{Evaluation of Risk and Resilience of the MBTA Green Rapid Transit System}

\author{
  Anil Kumar Gorthi \\
  \textit{Khoury College of Computer Sciences,} \\
  \textit{{Northeastern University}} \\
  \textit{gorthi.a@northeastern.edu}
}

\begin{document}
\maketitle

\begin{abstract}
The Transportation Systems Sector is one of the sixteen critical infrastructure sectors identified by the Cybersecurity and Infrastructure Security Agency (CISA) and plays a crucial role in ensuring public safety, economic stability, and national security. Within this sector, the mass transit and passenger rail subsectors are of particular importance due to their direct interaction with daily human mobility. The Massachusetts Bay Transportation Authority (MBTA) serves as the primary public transportation system within the Greater Boston Area and operates multiple subway, bus, commuter rail, ferry, and paratransit services. Among these, the Green Line is one of the oldest and most spatially complex rapid transit systems within the MBTA network. This paper presents a network-based risk and resilience assessment of the MBTA Green Line using graph theory, network metrics, and the Model-Based Risk Analysis (MBRA) tool. The Green Line network is simplified into a 17-node model derived from its original 70 stations, and multiple network metrics, including degree centrality, betweenness centrality, eigenvector centrality, spectral radius, node robustness, and blocking nodes, are calculated using Python-based analysis. Critical vulnerability is derived using the MBRA resiliency equation, and hypothetical random, targeted, and combined cyber-physical attack scenarios are constructed to evaluate system behavior under stress. The MBRA tool's results show that North Station, Government Center, Haymarket, Copley, and Kenmore are the most important nodes in the network. A fault tree analysis between Kenmore and Copley further demonstrates the impact of budget allocation on threat elimination and vulnerability reduction. Return on investment (ROI) is evaluated using MBTA financial data, and cybersecurity workforce requirements are mapped using the NICE framework. This work highlights key vulnerabilities within the Green Line network and provides actionable prevention, response, and workforce recommendations to improve resilience against cyber-physical threats.

\end{abstract}


\section{Introduction}
The Federal Government of the United States of America comprises fifteen departments within its Executive Branch, each led by a secretary holding a cabinet position in the White House. One of these executive departments is the Department of Homeland Security (DHS), which is divided into multiple sub-departments, agencies, and offices that fulfill various duties and responsibilities. One of the sub-departments of the DHS is the Cybersecurity and Infrastructure Security Agency (CISA), which is responsible for safeguarding the critical infrastructure of the United States. These critical infrastructures span across sixteen different sectors identified by CISA \cite{ref1, ref2}. One of these sectors is the Transportation Systems Sector.
This paper focuses on the Mass Transit and Passenger Rail sub-sector within the Transportation Systems Sector. The mass transit and passenger rail authority within the Commonwealth of Massachusetts is the Massachusetts Bay Transportation Authority (MBTA). Although the MBTA was formed as an independent organization, it became a division of the Massachusetts Department of Transportation (MassDOT) in 2009. The MBTA provides public transportation services via multiple modes, including commuter rail, subway, bus, and ferry. Additionally, there is a paratransit service known as “The RIDE.” The subway and bus transit systems are collectively referred to as the “T.” The MBTA offers services throughout the Greater Boston Area and certain regions beyond Boston and Massachusetts \cite{ref3}.

The subway system consists of the Green, Blue, Red, and Orange lines, along with a trolley train operating within the Red Line. The Green Line and the trolley train function as light-rail systems, whereas the Red, Orange, and Blue lines operate as heavy rail or metro systems. The Green Line consists of four sub-lines: B (from Boston College to Government Center), C (from Cleveland Circle to Government Center), D (from Riverside to Union Square), and E (from Heath Street to Medford/Tufts). These four Green Line branches, B, C, D, and E, form the primary area of focus of this study.

The Green Line is the oldest of the subway lines and serves one of the largest geographic areas within Greater Boston. It is also the most versatile and unique of the subway lines due to its varying physical characteristics. Portions of the Green Line run on the ground, on the road, sharing space with vehicular traffic, and underground. The B, C, and E lines operate on-road for significant portions of their routes and run underground from Kenmore, while the D line is the only branch that follows a fully separated, standalone right-of-way and does not share its path with vehicular traffic. A significant number of Green Line stations offer bus connections, ensuring seamless transfers, accessibility, and passenger comfort throughout the network. The subway vehicles operating on the Green Line are commonly referred to as “trolleys,” “trams,” or “trains.”

This paper examines the risk faced by, and the resilience of, each Green Line subway transit branch through a network-based modeling approach. It evaluates the structural and operational vulnerabilities of critical stops (nodes), identifies high-risk network components using graph-theoretic metrics, and applies the Model-Based Resilience Analysis (MBRA) tool to quantify network-level risk and resilience. The study further discusses system shortcomings, prevention and response strategies, risk mitigation approaches, resilience improvement measures for critical nodes, vulnerability elimination costs, and optimal budget allocation strategies. In addition, hypothetical cyber, physical, and combined attack scenarios are developed to understand the operational consequences of adversarial actions on key transit hubs within the MBTA Green Line network. 

This study reflects the operational, financial, and policy environment of the MBTA Green Line system as observed during late 2023, when the data, threat assumptions, and budgetary factors used for this analysis were collected and considered. Some of the security controls, staffing practices, fare validation mechanisms, and infrastructure modernization efforts discussed in this paper may have since evolved or are currently being implemented as part of ongoing system upgrades and resilience improvement initiatives. However, the network-based modeling framework, the risk and resilience assessment methodology, and the cyber-physical threat analysis approach presented in this study remain broadly applicable to present-day and future urban rail transit systems.

The key contributions of this paper are as follows: First, the MBTA Green Line network is modeled as a simplified 17-node graph derived from its full 70-station topology to enable tractable network risk analysis. Second, multiple network metrics, including degree centrality, betweenness centrality, eigenvector centrality, spectral radius, node robustness, and blocking nodes, are computed using Python-based analysis. Third, the MBRA tool is applied to derive network-level risk, critical vulnerability, and resilience characteristics. Fourth, hypothetical random, targeted, and combined cyber-physical attack scenarios are constructed for the Kenmore station to assess cascading system impacts. Fifth, a fault tree analysis is performed between Copley and Kenmore to evaluate the effect of budget allocation on vulnerability reduction and threat elimination. Finally, return on investment (ROI) and cybersecurity workforce requirements are evaluated using MBTA financial data and the NICE framework.

\section{Related Work}
Risk, resilience, and security analysis of transportation infrastructure have been areas of growing interest due to the increasing dependence of modern societies on highly interconnected transit systems and the rising frequency of cyber and physical threats. Prior research in this domain broadly spans modeling the resilience of transportation systems, network robustness analysis using graph theory, cyber-physical security of rail infrastructure, and policy-driven critical infrastructure protection frameworks.

Several studies have focused on the resilience of transportation networks using graph-theoretic approaches, where network connectivity, redundancy, and node criticality are quantified through metrics such as degree centrality, betweenness centrality, spectral radius, and node robustness \cite{ref12, ref13, ref14, ref15}. These works demonstrate that hubs with high betweenness and degree centrality often act as critical points of failure whose disruption leads to cascading operational breakdowns across the network \cite{ref16, ref17}. Network robustness and blocking node identification have been widely applied to power grids, airline networks, and urban road systems, establishing a mathematical foundation for understanding how infrastructure behaves under node or link removals. This body of work strongly motivates the application of similar modeling techniques to mass transit rail systems such as the MBTA Green Line.

From a cybersecurity and critical infrastructure standpoint, numerous studies and federal reports underscore that transportation systems are progressively evolving into cyber-physical entities. Modern rail systems depend heavily on Supervisory Control and Data Acquisition (SCADA) systems, digital signaling, automated fare collection, surveillance networks, and centralized operations control \cite{ref9, ref10, ref11}. Attacks such as the Stuxnet worm demonstrated the real-world feasibility of cyberattacks against industrial control systems, while ransomware campaigns such as WannaCry highlighted the indiscriminate nature of large-scale cyber threats affecting public services. These incidents illustrate how cyberattacks can translate directly into physical consequences, making rail transit systems a high-impact target class.

At the policy and governance level, the United States government, through DHS, CISA, and the Department of Transportation (DoT), has developed sector-wide risk management frameworks for the Transportation Systems Sector \cite{ref8}. The Transportation Systems Sector-Specific Plan and the Transportation Systems Sector Activities Progress Reports emphasize layered security, resilience engineering, information sharing, and joint coordination between public and private stakeholders. These reports establish the strategic necessity of integrating physical security, cybersecurity, and emergency response planning into a unified risk management architecture. The roles of the National Infrastructure Coordinating Center (NICC) and the National Cybersecurity and Communications Integration Center (NCCIC) have been formally defined within this ecosystem to enable real-time coordination and threat intelligence sharing across critical sectors.

Model-Based Risk Analysis (MBRA) has emerged as a structured framework for analyzing resilience in complex infrastructure systems \cite{ref18, ref19}. MBRA makes it possible to measure risk by using models of threats, vulnerabilities, consequences, prevention costs, and response costs. It further allows system-wide evaluation of how budget allocation influences vulnerability reduction and threat elimination through fault tree analysis and objective-based optimization. Prior applications of MBRA and similar resilience modeling tools have primarily focused on energy systems, industrial facilities, and logistics networks, with limited direct application to urban rail transit networks at the operational station level.

This paper differentiates itself from prior work in several ways. First, it applies a detailed network-based risk and resilience framework specifically to an urban light-rail transit system, namely the MBTA Green Line. Second, it combines graph-theoretic network metrics with MBRA-based resilience modeling and fault tree optimization to look at both the structural and financial sides of risk. Third, it explicitly develops cyber, physical, and combined attack scenarios at a major transit hub (Kenmore) to evaluate cascading operational effects. Finally, it goes beyond just technical modeling to include estimating return on investment and making recommendations for cybersecurity workers using the NICE framework. This connects engineering analysis with workforce development and policy issues.

\section{Network Characterization}
The Green Line is further divided into four sub-lines, namely B, C, D, and E \cite{ref6}, offering services from Boston College, Cleveland Circle, Riverside, and Heath Street, respectively. All these sub-lines of the green line together comprise a network of 70 stations spanning across 43 kilometers (27 miles) approximately. These 70 stations or stops within the B, C, D, and E green lines are lessened to 17 by merging multiple stops on a straight path. Within the B line, the stops from Boston College to Blandford Street (16 stops) are merged, and a direct link between Kenmore and Boston College is placed. The B line has the greatest number of stops after Kenmore. In a similar manner, the stops from Cleveland Circle are merged, and a link between Kenmore and Cleveland Circle is directly placed, without mentioning the other 13 stops between them. Coming to the D line, Brookline Village is mentioned after Kenmore, primarily to distinctly mention the D line network within the graph, also noting that Brookline Village is the only stop with bus connections on the D line and is frequently observed to have a higher passenger flow even during non-peak hours. The E line, unlike others, does not merge with the green line at Kenmore but instead at Copley. The network graph explicitly mentions a link between Copley and Heath Street, which merges the 11 stations between them. This arrangement lowers the total number of stops to 17. The network graph of all the green line transit rails is mentioned below \cite{ref7}.
\begin{figure}
    \centering
    \includegraphics[width=1\linewidth]{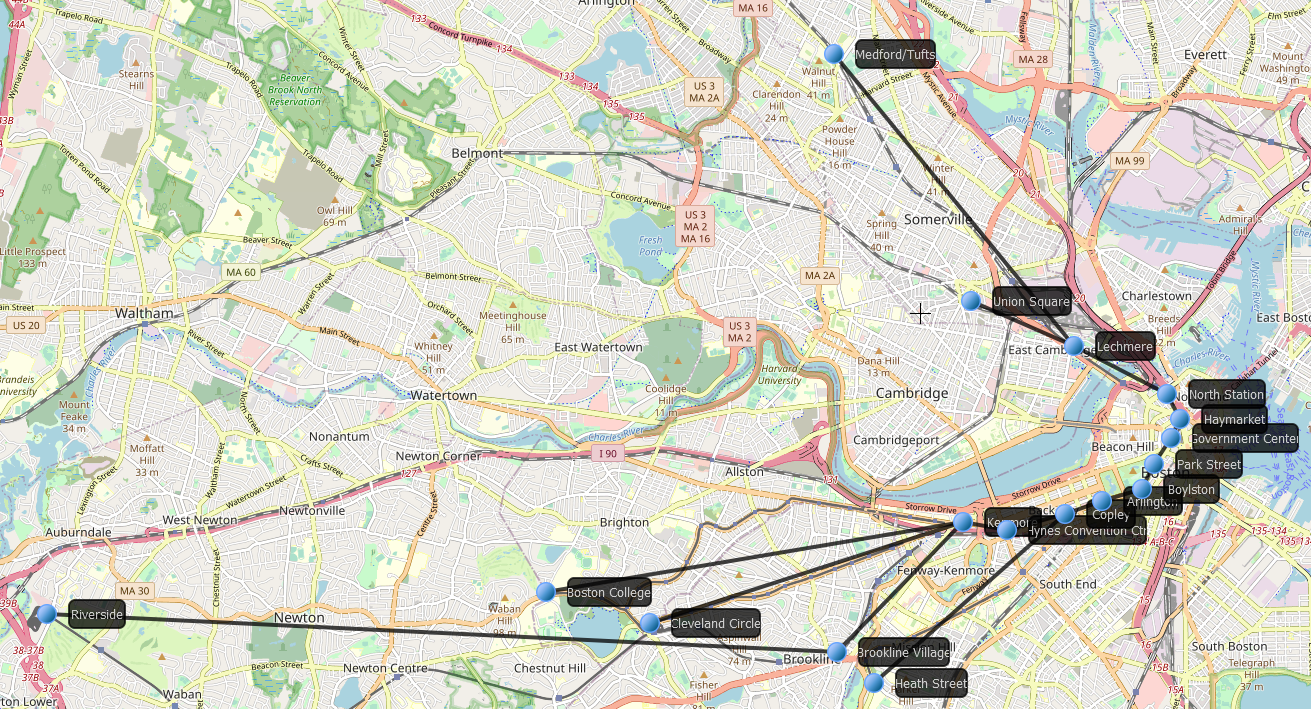}
    \caption{GPS Network Map}
    \label{fig:placeholder}
\end{figure}
\subsection{Nodes}
There are a total of seventeen nodes in this network, and each node within this network has its own importance and has a distinct contribution to the daily operation of the subway line. The outliers, or the end points or end nodes, of this network are Boston College (B), Cleveland Circle (C), Riverside (D), Government Center (B, C), Union Square (D) and Medford/Tufts (E). Although the Boston College (B) and Cleveland Circle (C) lines share the same route from Kenmore, they come to an end at Government Center (B, C), while the Riverside (D) line ends at Union Square (D) and the Heath Street (E) line ends at Medford/Tufts (E). All the lines share the same route starting from Copley, making Copley one of the critical nodes within this network. Kenmore is another critical node, which is a hub for the B, C, and D lines. All those lines join at Kenmore, making it a critical node with the highest degree. The vulnerability of each node is estimated based on an informed decision process, i.e., whether a station is underground, on-surface, or shares its path with vehicular traffic. Lines B, C, and E have their paths on-road, sharing a significant portion of their route with vehicular traffic. The vulnerability of each node has been estimated, considering location, average passenger flow, security, potential impact, and personal observation. Security is assessed by considering the presence of MBTA transit ambassadors or Transit Police personnel along with fare validation (i.e., in-train or at the entrance). All underground stations, along with some on-ground stations, require passengers to validate their fare before entering the station. Also, all stops that run on-road, along with some on-ground stations, have in-car fare validation (i.e., post-boarding). Riverside is the node considered to be the least vulnerable, with a vulnerability percentage of 50\%. It is considered so, as it is an on-ground station located far outside Boston, in Newton, MA, and requires fare validation to enter the station along with the presence of Transit Ambassadors. If a physical or cyberattack disables it, it will affect the network less than other nodes. 

Kenmore, an underground station to the west of Boston, is considered to have a vulnerability of 60\% as it is a hub for the B, C, and D lines with a moderate to high passenger flow during all the operational hours. The same vulnerability is shared by Hynes Convention Center, Copley, Arlington, Boylston, Union Square, and Medford/Tufts stations. Hynes Convention Center – the next stop after Kenmore for the B, C, and D lines - is similar to Kenmore in location (underground), flow of passengers, and security of station, hence the reason for considering the same vulnerability. Copley, Arlington, and Boylston are the order of stops when travelling eastwards in any of the green sub-lines. These three stops are similar in nature, so they are allotted the same vulnerability of 60\%. Union Square and Medford/Tufts also join the group of Kenmore, Hynes Convention Center, Copley, Arlington, and Boylston by sharing the same vulnerability (60\%), although they are outliers of the network. Their vulnerability of 60\% is due to their location (standalone and on-surface) and their unique characteristic of sharing their path with commuter rails. This makes them more vulnerable than Riverside, although they are outliers. Lechmere is the only node allotted a vulnerability of 70\% due to its unique characteristics. It is the only hub that is located over the ground level, and any attack on it will deny service to the Union Square (D) line to the northwest and the Medford/Tufts (E) line to the north. The Commuter Rail lines may also be affected in the event of a physical attack, cyberattack, or a combination of both; hence, Lechmere is allotted a 70\% vulnerability. 

Cleveland Circle, Boston College, and Park Street are allotted the same vulnerability of 80\%. It is so, as Cleveland Circle and Boston College are in important locations where there is a high population density, and they both are on-road stops, making them more vulnerable to physical attacks. Park Street, although being an underground stop, is as vulnerable as Boston College and Cleveland Circle because it is a common stop for the B, C, D, and E lines and for the red line too. It is located underground and close to the Government Center. Any physical attack or cyberattack will disable service north of Boston and impact the red line too. We assign a 90\% vulnerability to Government Center, Haymarket, North Station, Heath Street, and Brookline Village. Government Center, an underground station like Haymarket and North Station, is a terminus for the B and C lines, a hub for the D and E green lines, and the blue line. Any impact to this vulnerability will not only deny further service to the entire green line but will also significantly impact the blue line too. Haymarket is a hub for the D and E lines and the orange line. Although it has a constant presence of transit police personnel and transit ambassadors, it is highly vulnerable due to its location in downtown Boston. Any physical or cyberattack will have a severe impact on the operation of the subway and commuter rail lines, along with inconvenience to the surroundings or, worse, a catastrophe downtown. Brookline Village, an on-ground stop, is one of the highly vulnerable stops due to its location in the heart of the Town of Brookline, having a bus connection (route 66 Nubian Station to Harvard via Allston), and having no presence of MBTA personnel. This stop is one of the many which allow post-boarding fare validation So, entrance to the stop is not limited to passengers. Any type of attack on this stop will not only disrupt service for the D line but also may result in denial of bus service and have a significant impact on the neighboring community. Consequence, prevention costs, and response costs are allotted to each node after considering the characteristics of each node (degree and centralities).

\subsection{Links}
Line B from Boston College to Government Center shares it route with vehicles on Commonwealth Avenue, starting from the Blandford Street stop till the Boston College stop. Line C from Cleveland Circle to Government Center runs on Beacon Street from the stop on St Mary’s Street till the Cleveland Circle stop. Line E from Heath Street to Medford/Tufts runs on Huntington Avenue starting from the Northeastern University stop till Mission Park and proceeds into South Huntington Avenue starting from the Riverway stop till the Heath Street stop. Line D from Riverside to Union Square is a peculiar one, compared to the other lines (B, C, and E), as it shares no portion of its path with vehicular traffic. It has a separate, standalone, and right-of-way path in its entirety, which makes it rather special and a less vulnerable line than the others (i.e., B, C, and E). All the links are considered to have a vulnerability of 80\%, a consequence of \$8 million, a prevention cost of \$0.71 million, and a response cost of \$1.02 million (mentioned in the Appendix).

The network shows Kenmore directly connected to Boston College, Cleveland Circle, and Riverside. This is done to simplify the already complex transit network of the MBTA subway, easing estimation and calculation processes. The same type of connection between Copley and Heath Street and Lechmere and Medford/Tufts is shown to simplify the network. Using spreadsheet analysis and Python code, various calculations have been performed on this network. The constructed adjacency matrix (mentioned in the appendix) of the network along with the total number of nodes (17).

\subsection{Metrics}
Various network metrics (mentioned in the Appendix) that were calculated: Links/Node Degree, Network Degree, Average/Mean Degree, Spectral Radius, Link Robustness, Node Robustness, Blocking Nodes, links that can be removed, nodes that can be removed, and blocking nodes that cannot be removed. Link/Node Degree defines the number of links connected to a node, and the total of this is 32. Network degree is simply the maximum of link/node degree, which is 4 (degree of Kenmore). Average/mean degree is calculated by dividing total node degree (32) by total number of links (16), which is 2. Spectral radius can be used to estimate the connectivity of a network \cite{ref14}. It is calculated using the constructed adjacency matrix. The matrix is used to calculate eigenvalues of connection, which are then used to derive the spectral radius of this network, which is approximately 2.22. The Python code, along with the derived spectral radius, is mentioned below in the appendix. Along with Spectral Radius, Link Robustness, Node Robustness, Blocking Nodes, links that can be removed, nodes that can be removed, and blocking nodes that cannot be removed are also calculated (mentioned in the Appendix) using spreadsheet and Python code.

Link robustness is used to derive the resilience of a network when one or more links are removed. Here, link robustness is calculated as 0, by using the formula: $$[1-(2/Mean Degree)] $$ Node Robustness is used to evaluate the network’s withstanding capacity (i.e., resilience) when one or more nodes are removed. It is calculated as 54.89\% by using the formula $$[1-(1/Spectral Radius)]$$ Some nodes, when removed, significantly impact the network. Such nodes are denoted as blocking nodes. It is calculated as 45.11\% by using the formula [1/Spectral Radius]. The number of links that can be removed is 0 and is calculated by multiplying the total number of links and link robustness. The number of nodes that can be removed is 9 and is calculated by multiplying the total number of nodes with node robustness. The number of blocking nodes that cannot be removed is 8, which is calculated by considering This refers to the total number of links that include the specified blocking nodes. Along with these, degree centrality, betweenness centrality, and eigenvector centrality are also calculated using Python code. 

Degree centrality is a measure used to identify the nodes that are well-connected. Detailed values of each node are mentioned in the appendix. From Python calculations it can be understood that node 5 (Kenmore) has the highest degree centrality of 0.25, followed by nodes 7 (Copley) and 14 (Lechmere), which have a degree centrality of 0.1875. Calculation of degree centrality is based on node degree, and Kenmore, Copley, and Lechmere have node degrees of 4, 3, and 3, respectively.

\begin{figure}[H]
    \centering
    \includegraphics[width=0.75\linewidth]{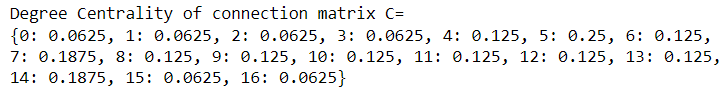}
    \caption{Degree Centrality}
    \label{fig:placeholder}
\end{figure}

Betweenness Centrality is a measure used to find the extent to which a node lies on the shortest path between other nodes. Node 7 (Copley) has the highest betweenness in this network with a value of 0.5750, followed by nodes 8, 9, 10, 11, 6, 5, 12, 13, 14, and 4 (Arlington, Boylston, Park Street, Government Center, Hynes Convention Center, Kenmore, Haymarket, North Station, Lechmere, and Brookline Village) in descending order, with values of 0.5333, 0.5250, 0.5000, 0.4583, 0.4583, 0.4416, 0.4000, 0.3250, 0.2416, and 0.1250, respectively.

\begin{figure}[H]
    \centering
    \includegraphics[width=0.75\linewidth]{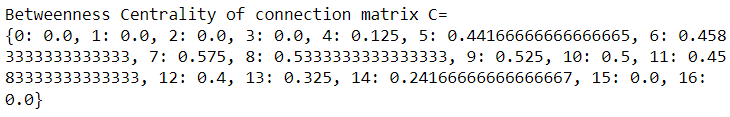}
    \caption{Betweenness Centrality}
    \label{fig:placeholder}
\end{figure}

Eigenvector centrality is based on the concept that connections to nodes with high degree contribute more to a node’s centrality, ultimately accounting for the quality and quantity of connections. Nodes with high eigenvector centrality not only indicate that they are well-connected but also connected to other well-connected nodes. Node 5 (Kenmore) has the highest eigenvector centrality, whereas node 16 (Medford/Tufts) has the lowest.

\begin{figure}[H]
    \centering
    \includegraphics[width=0.75\linewidth]{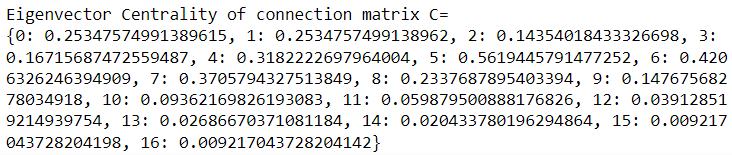}
    \caption{Eigenvector Centrality}
    \label{fig:placeholder}
\end{figure}

By calculation (mentioned in the Appendix) manually and using a Python script, the number of nodes that can be removed is 9. It seems that this number is derived by assuming continuity of service from Union Square and Medford/Tufts to Boston College, Cleveland Circle, Riverside, and Heath Street after removing the 9 nodes in between them except for Kenmore and Copley. It may be derived so, as all the nodes that can be removed are present consecutively with the same node degree, and the calculations are merely decreasing redundancy, ultimately optimizing the network.

The number of blocking nodes that cannot be removed is calculated as 8, which may be the outlier nodes (Boston College, Cleveland Circle, Riverside, and Heath Street) and the hubs joining them (Kenmore and Copley). Through calculations this number may be derived so, as the purpose of joining the outlier nodes and reducing the number of nodes between them would optimize the network. But this may not be practical to exactly implement the calculations due to one main reason – It thwarts the purpose of the subway line, i.e., public transportation and connectivity within the city.

\section{Resiliency Equation and Critical Vulnerability}

Using the MBRA tool, the resiliency equation and the critical vulnerability are calculated as explained. For the resiliency equation, the vulnerabilities of all the nodes and links are first set at 100\%, and the exponent value $q$ (i.e., the fractal dimension) is calculated by the MBRA tool itself using the chart palette. After this, the vulnerabilities of all nodes and links are set to 10\%, and the exponent value $q$ is calculated once again. Using the resiliency equation:
$$\log(q) = b + k\,\gamma\,\rho$$
The values of $b$ and $k$ are calculated (mentioned in the appendix). After deriving $b$ and $k$, they are substituted to find out the critical vulnerability by assuming $\log(q) = 0$. The calculated critical vulnerability for this network is 64\%.

The parameters used for this calculation are as follows:
\[
\text{Spectral Radius, } \rho = 2.22,\quad
\text{Vulnerability 1, } \gamma_1 = 10\%,\quad
\text{Vulnerability 2, } \gamma_2 = 90\%,\quad
q_1 = 1.86052,\quad
q_2 = 0.763993
\]

Substituting the first set of values into the resiliency equation,
\begin{align*}
\log(q_1) &= b + k\,\gamma_1\,\rho, \\
\log(1.86052) &= b + k(10\%)(2.22), \\
0.269 &= b + 0.22k \tag{1}
\end{align*}

Substituting the second set of values,
\begin{align*}
\log(q_2) &= b + k\,\gamma_2\,\rho, \\
\log(0.763993) &= b + k(90\%)(2.22), \\
-0.119 &= b + 1.998k \tag{2}
\end{align*}

By solving the above two equations, we get:
\[
-0.388 = 1.778k,
\]
which gives
\[
k = -0.218
\]

Substituting the value of $k$ into Equation (1),
\[
0.269 = b - 0.218 \times 0.22,
\]
\[
0.269 = b - 0.04796,
\]
\[
b = 0.31696
\]

Thus, the final derived values are
\[
k = -0.218,\quad b = 0.31696
\]

\begin{table}[H]
\centering
\renewcommand{\arraystretch}{1.3}
\begin{tabular}{|>{\centering\arraybackslash}p{0.45\textwidth}|>{\centering\arraybackslash}p{0.45\textwidth}|}
\hline
\textbf{$\log(q)$} & \textbf{$\gamma \rho$} \\
\hline
0.269  & 0.22  \\
\hline
-0.119 & 1.998 \\
\hline
\end{tabular}
\caption{Derived values of $\log(q)$ and $\gamma \rho$}
\end{table}

\begin{figure}[H]
    \centering
    \includegraphics[width=0.5\linewidth]{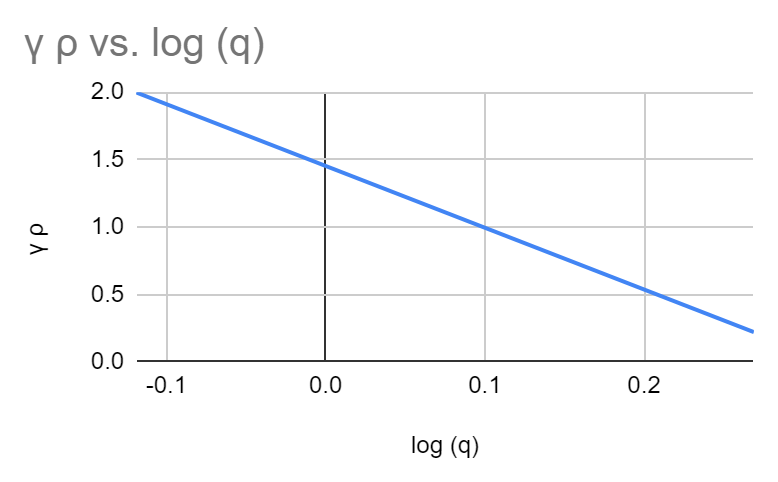}
    \caption{Calculating Critical Vulnerability}
    \label{fig:placeholder}
\end{figure}

Critical Vulnerability:

$$log (q) = 0,$$
$$0 = 0.31 - 0.218 \cdot \gamma \cdot 2.22,$$
$$-0.31 = -0.48396 \gamma,$$
$$\gamma = 0.640 \;\; \text{or} \;\; 64\%$$


\section{Random and Targeted Attacks}
Attacks on Critical Infrastructure and Key Resources (CIKR) are usually of two types: random and targeted. As the name suggests, random attacks are carried out without any specific targeting; hence, these are also known as indiscriminate attacks. The threat posed by these attacks is broad and unpredictable so, there cannot be a specific practice against a random attack. Implementing the latest and robust cybersecurity practices is the best solution to mitigate or prevent these attacks. Although there is not a specific target, the primary motive of the threat actor is to exploit the vulnerabilities of the desired CIKR. Improvisation to exploit vulnerabilities and cause the greatest damage by the threat actor may be witnessed during such attacks. Examples of random attacks are the WannaCry ransomware attack in 2017 and Distributed Denial-of-Service (DDoS) Attacks by the Mirai botnet in 2016. Although WannaCry spread globally and caused huge financial loss along with multiple deaths, it did not target any specific organization or industry. It is the same case with Mirai too. Although it launched multiple DDoS attacks and disrupted service on a large scale, it did not target any specific organization or sector.

Targeted attacks, unlike random attacks, do have a specific target. Although the motive and intent behind these attacks can be determined, defending from such attacks is more challenging, as it requires a multi-layer approach including an advanced threat detection system, employee training programs on social engineering techniques and active monitoring. A well-known example of this type of attacks is the Stuxnet worm attack in 2010. This malware was specifically designed to take over the Supervisory Control and Data Acquisition Systems of an Iranian nuclear facility by exploiting zero-day vulnerabilities.

\subsection{Hypothetical Random Attack Scenario at Kenmore}
Kenmore Station is in the Fenway community of Boston. Fenway is a popular location for sports fans, students, and concertgoers, so if there is an event in Fenway stadium (e.g., a Boston Red Sox match) or in any nearby clubs, the Fenway neighborhood will be filled with a great number of people, and the B, C, and D line of the MBTA subway will be packed with passengers travelling to Kenmore station. 
On a fine afternoon, Kenmore station is crowded with people. Students rush to catch their buses, commuters board their trains, and tourists navigate through Kenmore’s corridors. Amidst the crowd, there is a lone hacker with a computer, with their eyes glued to the screen and fingers on the keyboard. This hacker remotely launches an intricately and meticulously designed cyberattack on Kenmore Station’s infrastructure.

Initially, the hacker targets the electronic ticketing system (kiosks) beside the entrance of the station. Passengers are unable to purchase individual ride tickets or recharge their Charlie Cards, creating confusion and leading to long lines near those kiosks. Next, the fare validation machines at the entrance of the station are attacked and disrupted, preventing passengers from validating their Charlie Cards or ride tickets. These further fuel the chaos and confusion amongst travelers. With access to the station’s systems, the hacker then manipulates the timings of all the trains inbound and outbound from Kenmore.

As the ticketing system fails, the station’s public announcement system is then attacked, broadcasting a garbled warning. This infiltration on the sound system is carried out by the hacker to further intensify the confusion and chaos among passengers. Next, the hacker targets the digital boards, replacing advertisements with disturbing images and messages. This further advances the hacker’s goal to sow panic amongst travelers, creating a sense of vulnerability. Parallely, the hacker takes control over the surveillance system and takes the CCTVs offline by exploiting a vulnerability within the CCTV system. This would enable malicious threats actors working along with him to move around the station with great ease.

With the surveillance, PA, and ticketing systems compromised, the hacker now focuses on the control systems and takes over them. Most of the trains come to a halt outside Kenmore considering this cyberattack scenario, but a few running trains change tracks, run into other lines, and collide into other trains, escalating this situation to a catastrophe leading to mass casualties. From this series of events, it is evident that the hacker possesses the power to disrupt the entire transportation network originating eastwards and westwards from Kenmore.

As this attack situation keeps escalating, Kenmore’s emergency response system is triggered which sends automated alerts with codes to transit police, local police, state and federal law enforcement agencies, and emergency services (paramedic and fire brigades). However, this is anticipated by the hacker while they were taking over the station’s systems, so this emergency response system’s alerts are suppressed, and emergency communication channels are disrupted with the malware planted earlier.

Kenmore station descends into a state of absolute chaos with no one able to discern a plan of action or call for help. This random cyberattack on the Kenmore MBTA station will leave a lasting impact. This attack not only left the station’s and the B, C, and D lines’ infrastructure severely damaged but also tarnished the reputation of MBTA, raising concerns about the passenger, rail and station safety. This prompts the Department of Transportation to step and conduct investigations regarding the origin and cause of attack. National Transportation Safety Board (NTSB) will step in and conduct a thorough investigation and publish a report assessing the attack, risk, vulnerabilities, and resilience of Kenmore Station.

This attack serves as a stark reminder of the growing threat posed by cyberattacks on critical infrastructure. It highlights the need for vigilance, preparedness, and collaboration between local, state, and federal transportation authorities, cybersecurity professionals, and law enforcement agencies to safeguard public transportation systems, ultimately ensuring the safety of passengers.

\subsection{Hypothetical Targeted Attack Scenario at Kenmore}
The attack begins subtly with the hackers operating from a remote location and gaining unauthorized access to the station’s operational control systems. After successfully penetrating the station’s systems, they begin the first phase of the attack by manipulating the ticketing system, i.e., the kiosk at the entrance, which allows passengers to reload their Charlie cards or buy a ticket, through a vulnerability they found within it. This creates confusion amongst passengers and transit ambassadors posted there, leading to a state of psychological vulnerability.

In parallel, the hackers infiltrate Kenmore’s digital signage network, which allows them to morph advertisements into disturbing images and cryptic messages, gaining the attention of passengers and preying on their anxiety. Fear within passengers’ minds spread via the disturbing graphic images they posted through the signage boards. 

In the next phase, hackers launch a coordinated attack on the station’s PA system. They broadcast distorted messages, weaving tales of false emergencies and potential danger. They broadcast curated messages via the PA system, which would allow them to manipulate passengers. This leads to an escalation in chaos and panic, ultimately leaving passengers unable to act further.

Then, the hackers decide to further escalate the situation by seizing control of the train control system of Kenmore. They embed malware, which corrupts train data, leading to confusion within the drivers and Kenmore operations staff. Unable to procure MBTA data, trains inbound and outbound to Kenmore come to a halt, leaving passengers within the trains, in the dark underground tunnels, anxiously waiting to get out.

The hackers then target the station’s surveillance systems and take them offline, resulting in the blinding of the security staff’s digital eyes. As the station spirals down into chaos, they then unleash their final attack. They deploy a virus which would take down the station’s critical infrastructure, rendering them inoperable and useless. The virus also targets and disrupts emergency services systems, power, lighting, and communication channels, ultimately forcing everyone to panic in the now offline Kenmore station. This disruption of service cripples the operation of the B, C, and D lines beyond Kenmore, leaving everyone in a state of uncertainty.

Passengers panic within the dark station and force their way out of the station by breaking the entrance/exit barriers preventing them from leaving. After they find their way out, they reach emergency services and notify the law enforcement agencies. Transit, local, and state police respond immediately along with FBI and DOT personnel to assess the situation. The cause of attack, risks, vulnerabilities, and consequence are assessed.

This attack leads to lengthier and more thorough investigations by law enforcement agencies, state, and federal transportation departments, and CISA, unlike the random attack. Once the attack is determined to be targeted and planned, every aspect of the station and the subway line at large is questioned and reconsidered.

\subsection{Hypothetical Combination of Physical and Cyberattack at Kenmore}
This combined attack will first begin by hackers remotely infiltrating the station’s digital infrastructure, planting malware within those systems. This serves as a prelude to a more intricately planned attack. A team of attackers then enters Kenmore MBTA station without raising any suspicion. All their movements are planned and well-rehearsed. They then split into two groups, one targeting the station’s control center and the other targeting the platform.

In the control center, the attackers deploy the previously planted malware to its full potential. With a few clicks on their computer, they cripple the station’s critical infrastructure, including lighting, emergency system, communication system, and the train control system. Disruption of all these systems plunges the station into darkness and becomes a breeding ground for chaos and confusion.

Meanwhile, the second group of attackers on the platform implements the next phase of the attack. They carry out an assault wielding firearms and explosives to inflict maximum damage, resulting in widespread fear. Passengers’ panicked screams fill the air as passengers look for cover while being unsure of the next threat. 

The first group of attackers then proceeds with the next phase of the attack and disrupts the digital signage board. They display messages and propaganda on those boards, replacing advertisements. They also hack the public announcement system and broadcast messages, threats, and misinformation, fueling panic and anxiety among passengers. 

Being trapped in the dark and dangerous station, passengers must either navigate through the gunfire or seek refuge somewhere within the explosive-laced station. The attack is carried out by carefully identifying vulnerabilities, loopholes, and security lapses within the station and its systems. Once the motive of the attackers is fulfilled and escaped, or if they have been caught by police, extensive damage to the station and its infrastructure along with potential human casualties can be expected. This not only costs MBTA tens of millions of dollars but also tarnishes its reputation. This scale of attack leaves a mark within MBTA’s and Boston’s history. Emergency respondents and law enforcement personnel, along with state and federal investigators from various departments, work on assessing the Kenmore attack, which will lead to the identification of all the vulnerabilities within physical and cyber assets. The need for improvement in resilience, public awareness, and enhanced collaboration between public and private entities are identified and pointed out.

Considering most of the possible scenarios, the threat, vulnerability, consequence, prevention cost, and response cost have been allotted. Threat values throughout all the nodes and links are mentioned as 100\% because the MBTA should always remain under the assumption that an attack or a combination of attacks is impending. Vulnerability values are assigned to each node (stop) after considering the type of stop (on-ground, on-ground/open, or underground) and the location of the stop. For example, Riverside is located further away than all the other stops, and although it is on the ground, it is still protected through entry gateway systems, security personnel, and closed-circuit cameras – so vulnerability assigned is 50\%. Boston College and Cleveland Circle are both on-ground/open stops, which are in areas of medium-high traffic flow. Those stops are free to access by non-passenger so, they are more vulnerable than Riverside – hence, 80\% vulnerability for both. In the same manner, each node’s unique characteristics have been observed and the vulnerabilities assigned are through an informed/educated guess. Nodes of high importance and passenger numbers located in areas of high importance like Kenmore, Copley, Park Street, Government Center, Haymarket, and North Station are observed carefully and vulnerabilities are assigned based on their characteristics. Nodes lying at the ends of the network are in areas of less population density and although some of them are easy to gain access, they are less vulnerable because of the presence of CCTVs, security personnel and overall low passenger flow rate. The vulnerability of each link is mentioned the same as all links are equally vulnerable and have almost the same characteristics with very minute or negligible differences.

Consequence cost is assigned to each node solely based on passenger flow rate, location (if it is in an area of high importance), and overall population density in the vicinity of the stop. Prevention and response costs are derived from the MBTA’s Preliminary Itemized Budget for fiscal year 2024 \cite{ref4, ref5}. The estimated source of funding for prevention and response activities is from Materials, Services, Utilities, and Insurance mentioned under “Operating Expenses.” The total budget for all the green line trams is estimated to be \$150 million, which is further divided into two unequal halves: prevention and response budget. From the total, 60\% (\$90 million) is allocated as response cost and the remaining 40\% (\$60 million) as prevention cost. Nodes of higher importance like Kenmore, Copley, Park Street, Government Center, Haymarket, and North Station are assigned greater funding (50\%) — greater funding (\$4.5 and \$3 million each for response and prevention, respectively); nodes of medium-low importance like Brookline Village, Hynes Convention Center, Arlington, Boylston, and Lechmere are assigned 30\% of the prevention and response costs (\$3.24 and \$2.16 million each, respectively); and nodes lying at the ends (remaining nodes) of the network are assigned the lowest (20\%) funding of \$1.8 and \$1.2 million each for response and prevention, respectively.

\section{Formal Threat Model}
The threat model for this study is developed based on the hypothetical random, targeted, and combined cyber-physical attack scenarios constructed for the Kenmore MBTA station. Kenmore is selected as the primary focus for the threat model due to its high degree, centrality, and operational importance as a hub for the B, C, and D Green Line branches. Since all three branches converge at this station, any disruption at Kenmore has the potential to propagate both eastwards and westwards across the Green Line network.

\subsection{Adversary Model}
The adversary considered in this threat model includes both low-sophistication and high-sophistication threat actors. For random attacks, the adversary is assumed to be opportunistic with no specific targeting strategy and limited situational awareness. Such an adversary primarily exploits exposed vulnerabilities within public-facing systems such as electronic ticketing kiosks, fare validation machines, public announcement systems, digital signage boards, and surveillance infrastructure. The primary objective of the adversary in this case is disruption, confusion, and panic rather than strategic long-term damage. For targeted attacks, the adversary is assumed to be highly skilled, well-resourced, and motivated by deliberate intent. This adversary possesses advanced capabilities in cyber intrusion, malware deployment, control system manipulation, and coordinated execution. The objective in this case is not limited to service disruption alone but extends to reputational damage, infrastructure degradation, and large-scale operational paralysis. For combined cyber-physical attacks, the adversary is modeled as a coordinated group possessing both cyber capabilities and physical attack resources. This adversary is assumed to have the ability to infiltrate digital systems remotely, plant persistent malware, and then execute a synchronized physical assault within the station environment.

\subsection{Attack Surface}
The attack surface at Kenmore Station is composed of multiple cyber and physical components that are interconnected and operationally interdependent. On the cyber side, the attack surface includes electronic ticketing kiosks, fare validation terminals, public announcement systems, digital signage boards, CCTV surveillance systems, centralized control systems, and train control systems. These components are directly exposed to passenger interaction and network connectivity, making them attractive targets for exploitation. On the physical side, the attack surface includes station entry and exit points, platforms, underground corridors, power distribution systems, lighting, emergency communication infrastructure, and physical access to control rooms. Physical exposure is amplified due to high passenger density, underground confinement, and the public accessibility of station infrastructure. Because of the cyber-physical coupling of modern transit systems, an attack on digital assets can directly influence physical operations such as train movement, lighting, emergency response, and passenger evacuation.

\subsection{Impact Categories}
The impacts of a successful attack are categorized into four primary domains: operational, safety, financial, and societal. From an operational perspective, attacks can lead to complete service denial at Kenmore and cascading service disruption across the B, C, and D lines. Train scheduling, routing, and communications may become unreliable or inoperable. From a safety perspective, cyber-physical manipulation of train control systems, emergency systems, and lighting introduces the potential for mass casualties through derailments, collisions, stampedes, and delayed emergency response. From a financial perspective, the MBTA may incur costs associated with infrastructure repair, emergency response, legal liability, lost fare revenue, and long-term capital replacement. From a societal perspective, public trust in the safety and reliability of mass transit may be significantly eroded. Reputational damage to the MBTA and public fear regarding transportation security can have long-term consequences beyond the immediate physical damage.

\section{Limitations}
This study is subject to several important limitations, which must be acknowledged when interpreting the results. First, the Green Line network is simplified from its full 70-station topology into a 17-node model by merging multiple intermediate stops. While this simplification enables tractable network analysis and clearer visualization of structural relationships, it does not capture every operational nuance present in the full-scale system. Second, the vulnerability, threat, and consequence values assigned to each node and link are based on informed estimation using station location, passenger flow, fare validation type, underground versus on-surface characteristics, and personal observation. These values are not derived from classified threat intelligence or proprietary MBTA security data and therefore should be interpreted as approximations rather than absolute measurements. Third, the hypothetical attack scenarios developed for Kenmore Station are constructed to represent plausible cyber, physical, and combined attack pathways, but they are not intended to model any specific real-world adversary group or confirmed threat campaign. Fourth, the MBRA tool requires several modeling assumptions related to vulnerability scaling, budget allocation, and elimination cost behavior. While these assumptions are consistent with standard resilience modeling practices, they inevitably influence the numerical outcomes of the fault tree and ROI analysis. Finally, this study does not incorporate real-time sensor data, classified SCADA architecture details, live signaling protocols, or internal MBTA cybersecurity telemetry. As a result, the analysis reflects a macro-level resilience and risk perspective rather than a fine-grained operational cybersecurity audit.
Despite these limitations, the modeling approach presented in this study provides a structured and repeatable framework for evaluating risk and resilience in urban light-rail transit systems.

\section{Ethical and Public Safety Impact}
Public transportation systems directly interact with millions of passengers on a daily basis, and therefore any assessment of risk, vulnerability, and resilience must also account for ethical responsibility and public safety considerations. Unlike purely digital systems, failures within mass transit infrastructure can result in immediate physical harm, loss of life, and large-scale social disruption. As a result, studies such as this one carry significant ethical weight, particularly when analyzing cyber-physical attack scenarios and infrastructure vulnerabilities.
One of the primary ethical responsibilities associated with this research is the careful handling of sensitive infrastructure information. While this study identifies critical nodes, operational vulnerabilities, and potential attack surfaces within the MBTA Green Line network, it does so at a macro level without exposing classified system architectures, proprietary control system configurations, or exploitable technical details. This balance between transparency and responsible disclosure is essential to ensure that the research contributes to improved security without unintentionally enabling malicious activity.

From a public safety perspective, the outcomes of this study emphasize the importance of proactive risk mitigation rather than reactive emergency response. The identification of high-risk nodes such as Kenmore, Copley, Government Center, Haymarket, and North Station highlights locations where enhanced security controls, faster response mechanisms, and greater situational awareness can significantly reduce the probability of mass-casualty events. The prevention and response strategies discussed in this paper directly align with the ethical obligation of transit authorities to prioritize passenger safety over cost efficiency alone.

The hypothetical attack scenarios presented in this study are not intended to create fear or speculation but rather to illustrate how cyber and physical threats can cascade through interconnected transit systems if vulnerabilities are not adequately addressed. These scenarios serve as controlled analytical tools to evaluate system behavior under stress and to inform better design of security controls, emergency response protocols, and workforce preparedness.

Another ethical dimension of this work lies in resource allocation and equity. Security investments, workforce expansion, and infrastructure upgrades should not be concentrated solely in affluent or high-visibility locations at the expense of lower-income or peripheral communities. Many of the Green Line stops that rely on post-boarding fare validation and limited personnel presence serve dense residential neighborhoods. Ensuring that these locations receive adequate protection is essential to maintaining equitable public safety standards across the transit network.

Finally, this study reinforces the ethical obligation of continuous improvement in cybersecurity awareness, employee training, and public communication. Transit employees operate at the front lines of public safety and must be equipped with the knowledge and tools to identify early warning signs of cyber or physical threats. Similarly, passengers must be able to trust that the systems they rely on daily are actively monitored, well-maintained, and protected using modern security practices.
Overall, the ethical and public safety implications of this research underscore the need for sustained investment in resilience, transparency in risk communication, responsible disclosure of vulnerabilities, and an unwavering commitment to protecting human life within public transportation systems.

\section{Using MBRA to Identify Risks and Critical Nodes}
After mentioning consequence cost, prevention cost and response cost on the MBRA tool, we can calculate final risk by tapping on the calculate button within the Calculate section of the Tool Palette. The final/calculated risk for the entire network is 230.20. In the order of greatest to least, North Station (Risk = 18.00), Government Center (Risk = 15.30), Haymarket (Risk = 14.40), Copley (Risk = 9.00), Kenmore (Risk = 6.00), Park Street (Risk = 12.00), Lechmere (Risk = 6.30), Brookline Village (Risk = 9.00), Boylston (Risk = 7.80), and Arlington (Risk = 7.20). The graph regarding this is mentioned in appendix (figure 8). All the nodes in the figure are ranked according to “Degree” weight selected in the tool palette. According to the MBRA tool, the node with high criticality is North Station. This seems plausible, as North Station is one of the major train, bus, and tram of the MBTA. Amtrak, T-Commuter Rail, Green and Orange subway lines and bus lines operate from this location. With this, it is understandable that this stop has a high passenger flow and is also one of the most famous commercial sports arenas in Greater Boston (TD Garden located in the same building above the commuter rail station), which opens this stop for greater risk than any other. So, the critical nodes in this network are North Station, Government Center, Haymarket, Copley, Kenmore, Park Street, Lechmere, Brookline Village, Boylston, and Arlington.

\section{Prevention Controls in Risk Mitigation and Resilience Improvement}
There are various ways through which risk faced by a node and the overall risk faced by the entire network can be minimized through a robust prevention control, risk mitigation, and resilience improvement strategy. Starting with passenger safety measures, emergency communication systems need to be overhauled and updated if needed. This can be used to broadcast emergency messages to passengers so they may either defer their travel or make an alternate travel plan, putting forth their safety at the top. There is also a need for personnel at large hubs like Kenmore, Copley, Haymarket, North Station, etc., who can handle emergency evacuation effectively and can guide passengers to safety. Currently, there are personnel at said hubs but the number needs to be increased to manage passengers during unforeseen circumstances. Most on-ground stops on the green line are open-access, which significantly increases the risk of such stops, without any protective physical barriers to prevent non-passengers from entering the stop. This prevention method is for the trams going on roads and sharing their path with vehicular traffic—dedicated traffic signals for trams and newer traffic signal time periods for tram and non-tram traffic would not only help trams avoid any potential accident scenario but also help in reducing loss of lives and property. There is also a need for an increase in surveillance measures in open stations through both physical and digital solutions. Patrol personnel on-ground along with an increase in the number of CCTVs, would help secure such stops for all passengers. Most importantly, employees need to be trained and be made aware of the latest cybersecurity practices to safeguard the equipment and machinery in case of a threat or attack. Every node (stop) needs to be under the assumption that it is going to be a target of a future combined physical and cyberattack. This will help the staff to train under all possible major threat scenarios like natural disaster, cyberattack, and physical attack (including a terrorist attack). This policy would help prevent many of the attacks of all forms.
\section{Response Controls in Risk Mitigation and Resilience Improvement}
There needs to be a response strategy for a particular kind of attack so that the staff can be prepared to handle all the passengers safely during unfortunate circumstances. Generally, for accidents or a natural disaster, emergency response plans need to be put in place specifically for each station so that there can be a quick response from staff and passengers, ultimately reducing or eliminating loss of lives. Adopting the Incident Command System (ICS) so that there can be a swift and efficient coordination between various staff members and utilization of required resources during an incident. Adopting and implementing emergency communication protocols to relay information to staff and passengers in case of a physical or cyber attack and a natural disaster. Not only employees, but passenger awareness needs to increased regarding emergency communication protocols and emergency equipment. In case of a terrorist attack, there needs to an emergency evacuation protocol and services put in place so that passengers can be safely escorted away from threat. While all these strategies are implemented, there is also a need for establishing recovery plans and teams of personnel who can analyze and help catalyse the recovery process. Although updates to plans and strategies are required as per the ever-evolving landscape of threats, monitoring is one of the key strategies which can ensure effective implementation of all the required protocols and strategies. For example, an implementation of the above strategies in a critical node like North Station would help in ensuring passenger safety and minimizing damages and injuries. Communication and effective sharing of information with law enforcement agencies is another key aspect of improving response time for mitigating risk and improving resilience of a node (stop). Focus on critical nodes should be greater to ensure passenger safety and safeguard critical equipment. Links with high betweenness tend to be the ones with greater risk, so deployment of more personnel and strengthening access controls would bolster the security of the node, minimizing risk.

In the D line, from Fenway towards Riverside, all the stops are on the ground but standalone, i.e., do not share their path with vehicular traffic. But these stops are vulnerable, as they all have post-boarding fare validation, which opens the stops for non-passengers too. This increases the vulnerability of each stop, as literally anyone can enter and meander through a stop. Unlike underground stops like Kenmore, Copley, or Park Street, these stops do not have any MBTA personnel present in the station. To increase security and resilience of such stops, fare validation barriers need to be put up along with the presence of transit ambassadors and transit police within each stop.

In B, C, and E lines, a significant number of stops are on-road, some with platforms and some without platforms. Although the stops that do not have a platform have multiple factors influencing their vulnerability and changes cannot be made to those stops, the security of the ones that have a platform can be strengthened by the presence of transit police personnel. If an unfortunate event is impending, the presence of transit police personnel can help in initiating emergency response as soon as possible.
Hiring new personnel to take care of security would help reduce the risk for each node. Since this is a recurring expenditure and assuming that a single staff member’s salary would be \$70,000 per annum and the required number of personnel are 20 to bolster the security of a node, the total expenditure would be \$1.4 million. Strengthening access points of a node by replacing all the old and outdated passenger access management terminals with newer ones, assuming each one costs \$10,000 and considering the node (stop) has two entrances, requiring a total of 10 terminals (5 for each entrance), would lead to a total cost of \$100,000. If the node has a greater number of entrances and is large in area, then it would require a greater number of terminals to completely secure the entrances. If an emergency evacuation facility is set up at a node, it would require the purchase of a fleet of mini-buses which could safely carry passengers away from the threat. Assuming the cost of each bus would be \$500,000 and 20 of such buses are required for a node, then the total would be \$10,000,000. These are some of the risk mitigation and resilience improvement strategies and their budget for one node, totalling \$11.5 million for one node.

\section{Fault Tree Analysis Using MBRA Tool}
The fault tree analysis is carried out between Copley and Kenmore, two major hubs of the green line. The threats considered are bomb and SCADA, i.e., one physical attack and one cyberattack for each node. The total budget for both the nodes is \$10 million, with the elimination cost for Bomb at both Copley and Kenmore set as \$3 million and the elimination cost of SCADA at Copley and Kenmore set as \$2 million. The consequences for Bomb and SCADA at Copley and Kenmore are \$5 million and \$10 million, respectively. Initially when the budget is set to \$0, the vulnerability does not reduce, and the risk remains the same too. As the budget is allocated for the elimination of the threats is increased to \$5 million, it is observed that the threat faced by Bomb significantly reduces, whereas the SCADA threat reduces only by a slight amount. At this point, the overall vulnerability reduces to 32.75\% from 50.23\%, and the risk is also reduced from 4.90 to 2.64. 

When the budget is increased to \$10 million, i.e., to the maximum, it is observed that both Bomb and SCADA threats have reduced to the same level at both Copley and Kenmore. The overall risk fell to 1.50 from 4.90 and the vulnerability fell from 50.23\% to 18.55\%. From this, it is evident that the initial effect of the elimination cost can be felt by the reduction of threat from Bomb and then SCADA. But, at the end when the maximum budget is allocated for elimination, although the threats are not eliminated, they are significantly reduced to the same level. (figures mentioned in appendix).

\begin{figure}[H]
    \centering
    \includegraphics[width=1\linewidth]{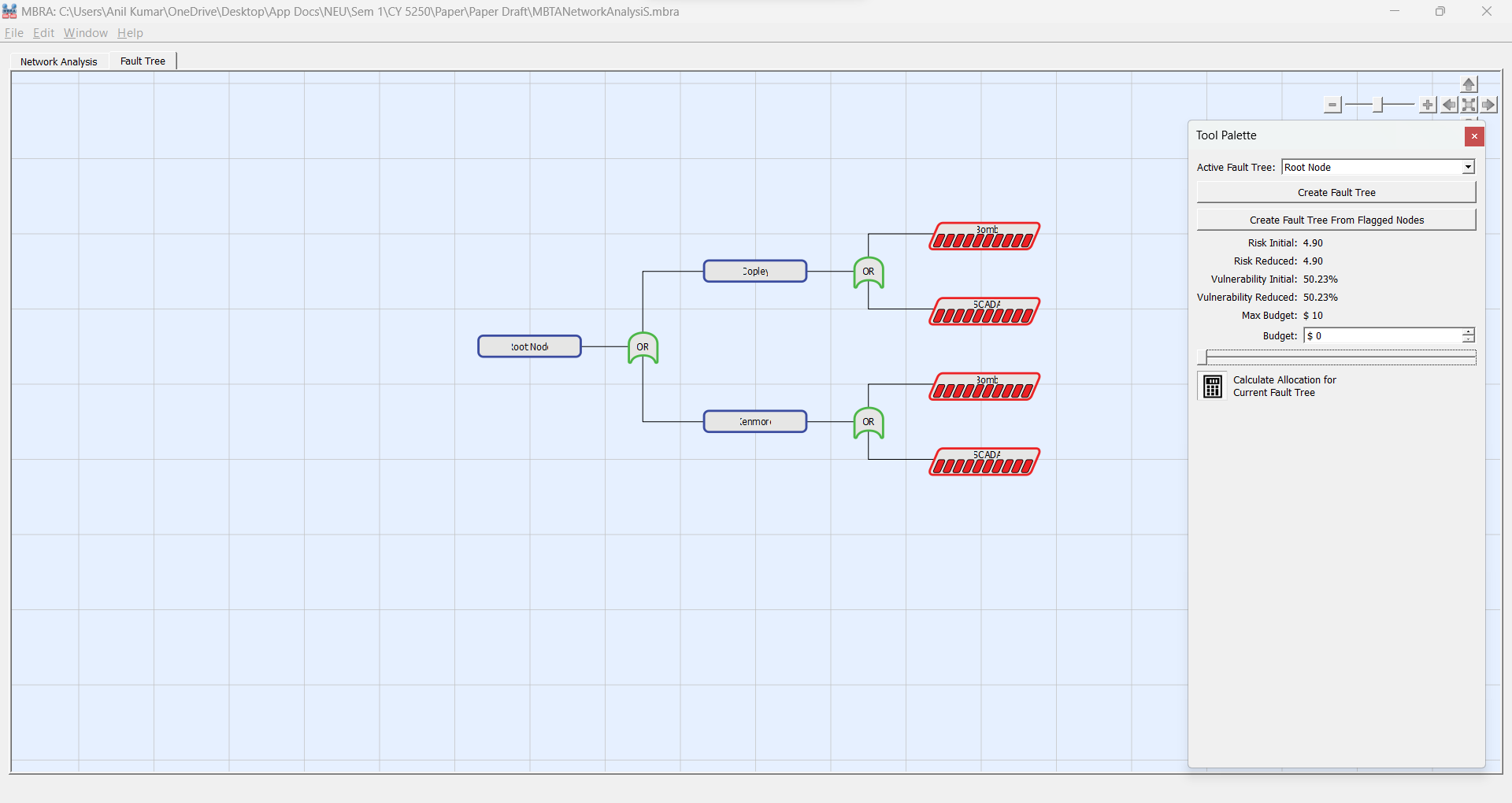}
    \caption{Fault Tree with no budget utilized}
    \label{fig:placeholder}
\end{figure}

\begin{figure}[H]
    \centering
    \includegraphics[width=1\linewidth]{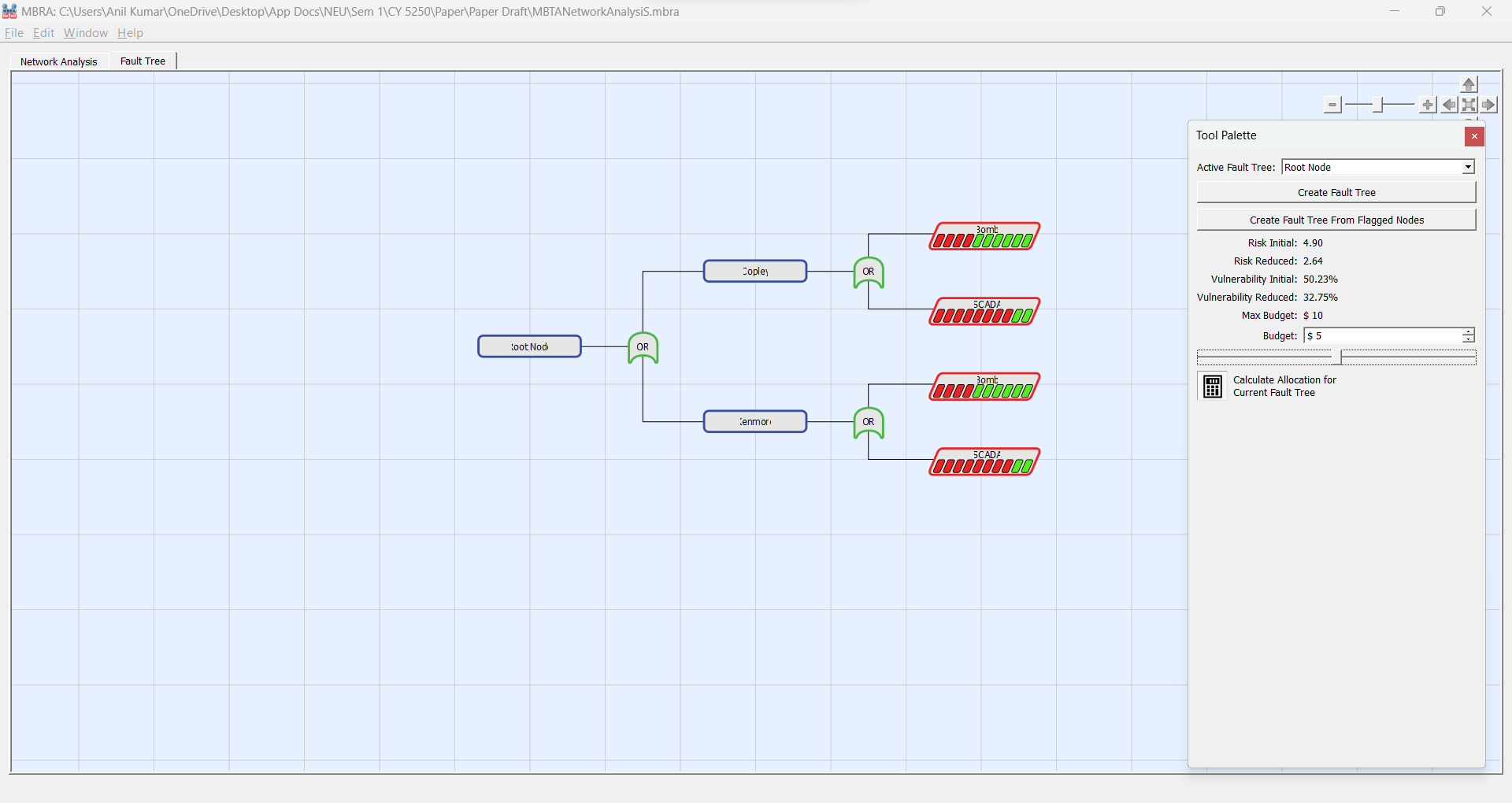}
    \caption{Fault Tree with half budget utilized}
    \label{fig:placeholder}
\end{figure}

\begin{figure}[H]
    \centering
    \includegraphics[width=1\linewidth]{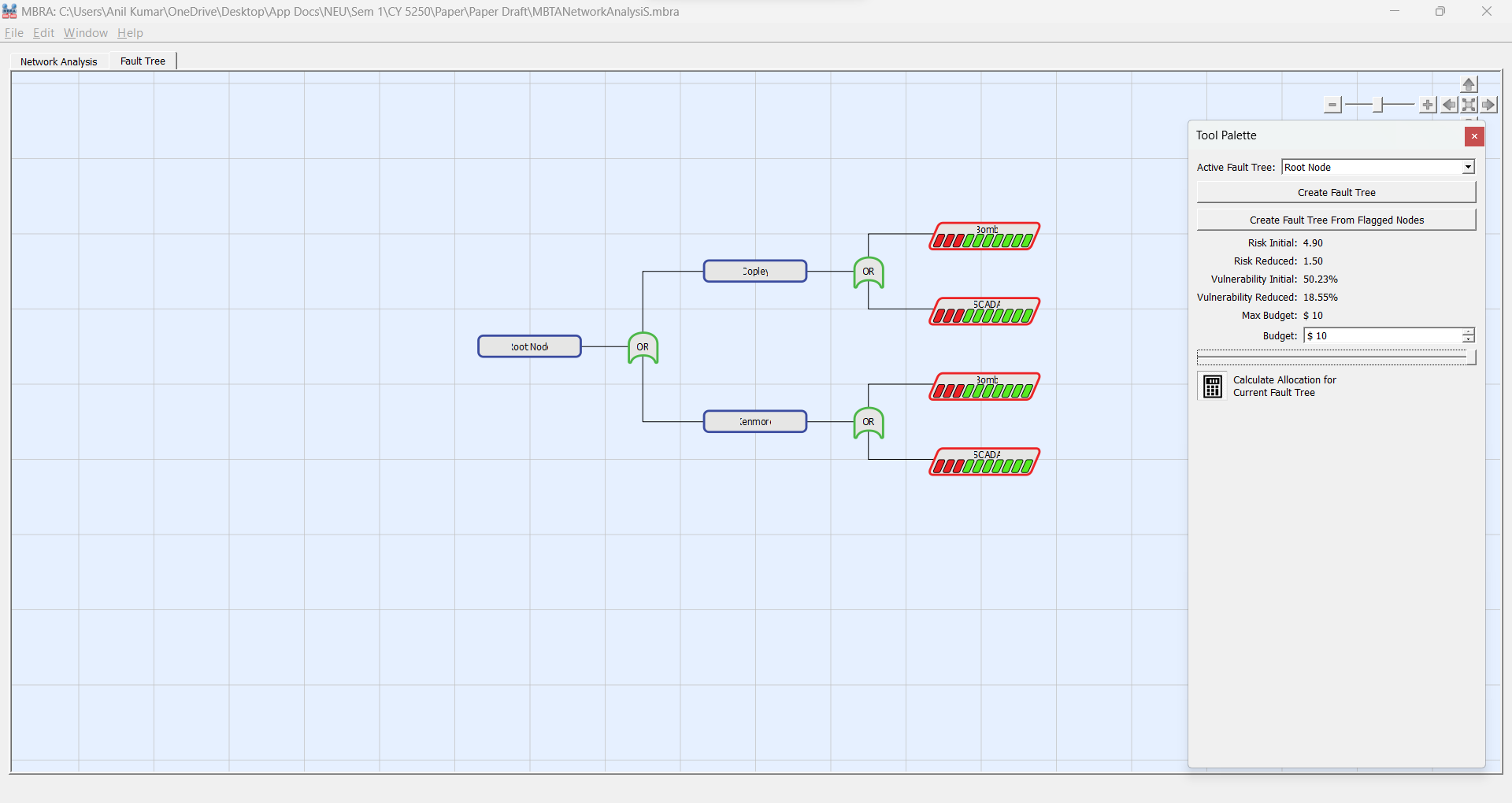}
    \caption{Fault Tree with full budget utilized}
    \label{fig:placeholder}
\end{figure}

\section{Role of NICC and NCCIC}
The National Infrastructure Coordinating Center (NICC) is the main coordinating body on critical infrastructure protection and resilience improvement. This organization is responsible for facilitating information sharing and collaboration with other CISA sectors, including transportation, by addressing both physical threats and cyber threats. NICC collaborates with local, tribal, state, territorial, and federal organizations or agencies to manage responses to incidents effectively \cite{ref8}. The NICC provides information related threats, vulnerabilities, and possible attacks by analyzing and monitoring situations related to critical infrastructure. NICC’s information is crucial when it comes to effective decision-making. The National Cybersecurity and Communications Integration Center (NCCIC) is a key division of CISA (Cybersecurity and Infrastructure Security Agency) and plays a key role in increasing resilience of critical infrastructure across various sectors including transportation. It mainly focuses on the identification, response, and mitigation of risks from the cybersecurity front. It is a hub of threat intelligence with public and private sector partners, so it has access to real-time data, which helps it analyze information related to emerging threats, vulnerabilities, and best practices to enhance the resilience of a sector. In the aftermath of an unfortunate event, for example, a cyberattack on an MBTA stop which led to passenger and tram information manipulation and chaos, NCCIC helps in coordinating efforts to rectify the situation and restore normalcy as soon as possible. Also, NCCIC provides cybersecurity-related training to entities (in this example, the staff of MBTA) to increase awareness and skills of the employees. Both NICC and NCCIC work together to secure various critical infrastructures within the Transportation Systems Sector. They actively work to secure seaports, airports, train stations, and bus transit systems with numerous private and public sector partners to increase the resilience of this sector.

\section{Return on Investment and Possible Funding Sources}

$$\text{Risk Initial } = 230.20$$
$$\text{Risk Final } \text{(calculated) } = 128.08$$
$$\text{Expenditure} = \text{Prevention} + \text{Response Costs} = \$10 \text{ million}$$
$$\text{Return on Investment} = (\text{Risk} \text{ Initial} - \text{Risk} \text{ Final})/ \text{Expenditure}$$
$$= (230.20 - 128.08)/10$$
$$= \$10.212 \text{ million}$$

Return on investment can be calculated as mentioned above, i.e., the difference between initial and final risk divided by expenditure.

MBTA has two sources of revenue: operating revenues and non-operating revenues \cite{ref3, ref4, ref5}. Operating revenues include fares of all modes which total \$418.5 million, and own-source revenues which includes advertising, parking, real estate, and other operating revenues totalling to \$82.4 million, which brings the entire operating revenue to a grand total of \$500.9 million. The non-operating revenues consists of dedicated sales tax of \$1463.5 million, dedicated local assessments of \$188.4 million, other income of \$22.9 million, federal funds of \$31.4 million, and state-contract assistance of \$441.4 million. This totals the non-operating revenue to a grand total of \$2147.4 million, bringing the entire revenue of the MBTA to a final total of \$2648.3 million.

\begin{figure}[H]
    \centering
    \includegraphics[width=0.5\linewidth]{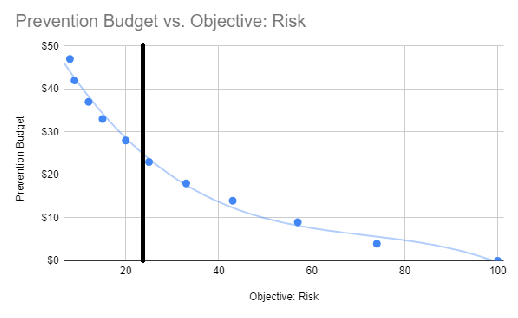}
    \caption{ROI declines as vulnerability reduction funds increase due to diminishing returns}
    \label{fig:placeholder}
\end{figure}

In the section “Response Controls in Risk Mitigation and Resilience Improvement,” the need for placing fare validation barriers was discussed as one of the solutions to improve resilience and security of the stops within the D line. This would not only boost the security of a stop but also help in generating revenue from every passenger. The green line, which covers almost 50\% of the distance within the MBTA’s rapid transit system, relies mostly on post-boarding fare validation of passengers. But this can be achieved when passengers board through the front door of each car. Since not all passengers board in that manner, there is a significant loss in revenue every day at every stop. To implement higher security standards, hire more security personnel, and update old digital infrastructure, MBTA needs to ensure 100\% fare validation, i.e., every passenger needs to contribute their share, so MBTA can invest to reduce risk and potential vulnerabilities.

\section{Proposed Cybersecurity Improvements (Call to Action)}
The on-ground stops in the B, C, and E lines are open to physical attacks in multiple ways. Although fare validation barriers at the entrance of each station would not be completely effective, the presence of personnel (either Transit Ambassadors or Transit Police) will make a significant improvement to the security of each stop. The presence of personnel will not only make threat actors question themselves, but it will also help in significantly reducing the response time for transit and local police along with paramedics and fire brigades. Ultimately, the risk and vulnerability of each stop is lessened while resilience is improved.
The D line, which has standalone and right-of-way stops unlike all the other sub-lines of the green line, requires the presence of MBTA personnel (either Transit Ambassadors or Transit Police) along with fare validation barriers at the entrances/exits of every stop. This will significantly boost the security and resilience while reducing the risk and vulnerability of each stop.

\section{Cybersecurity Workforce Recommendations}
To boost resilience and security and reduce risk and vulnerability of on-ground stops within the B, C, D, and E lines, the presence of Transit Ambassadors to ensure fare validation and provide passenger support, along with the presence of Transit Police personnel to boost security of those stops, is required. As per the NICE framework published by NIST, an index of the tasks and Knowledge, Skills, and Abilities (KSAs) of the required personnel to be hired for one stop (node) is mentioned below.

\subsection{Human Capital Index}

For every stop (node), there is a requirement to hire at least one transit ambassador and one transit police person. So, a total of two people per stop. The title as per NICE framework along with the required KSAs and tasks to be performed, is mentioned below.

\begin{table}[H]
\centering

\textbf{Title: Security Control Assessor (SP-RSK-002)} \\[8pt]

\renewcommand{\arraystretch}{1.2}
\begin{tabular}{p{0.18\textwidth} p{0.75\textwidth}}
\toprule
\textbf{KSA ID} & \textbf{Description} \\
\midrule
K0267 & Knowledge of relevant laws, policies, procedures, or governance related to critical infrastructure. \\
S0006 & Skill in applying confidentiality, integrity, and availability principles. \\
S0027 & Skill in determining how a security system should work and how changes in conditions, operations, or the environment will affect these outcomes. \\
S0034 & Skill in discerning the protection needs (i.e., security controls) of information systems and networks. \\
S0086 & Skill in evaluating the trustworthiness of the supplier and/or product. \\
\bottomrule
\end{tabular}

\vspace{6pt}

\caption{KSAs for the role of Security Control Assessor (SP-RSK-002)}
\end{table}
\begin{table}[H]
\centering

\renewcommand{\arraystretch}{1.2}
\begin{tabular}{p{0.18\textwidth} p{0.75\textwidth}}
\toprule
\textbf{Task ID} & \textbf{Task} \\
\midrule
T0079 & Develop specifications to ensure risk, compliance, and assurance efforts conform with security, resilience, and dependability requirements at the software application, system, and network environment level. \\
T0221 & Review authorization and assurance documents to confirm that the level of risk is within acceptable limits for each software application, system, and network. \\
\bottomrule
\end{tabular}

\vspace{6pt}

\caption{List of tasks for the role of Security Control Advisor}
\end{table}

\section{Conclusion}
An examination of computations, attribute traits, and theoretical scenarios at all transit stops reveals numerous vulnerabilities that can significantly heighten risk levels. The author acknowledges the theoretical possibility of complete risk elimination, but it may be improbable to achieve this result in practice. In conclusion, the D line stations exhibit a lower risk profile compared to the B, C, and E lines, as demonstrated by the D line's minimal calculated risk, supported by personal observation and analyses conducted using Model Based Risk Analysis and calculation via Python code. Optimizing passenger fare validation on the green lines (B, C, D, and E) is essential for the Massachusetts Bay Transportation Authority (MBTA) to generate revenue for infrastructure improvements and transit-related studies.

\section*{Acknowledgments}
This work originated as a course project in CY 5250 (Decision Making in Critical Infrastructure) at Northeastern University (Boston, MA, United States) in November 2023.

\bibliographystyle{unsrt}
\bibliography{references}

\appendix
\section{Appendix}

\begin{figure}[H]
    \centering
    \includegraphics[width=1\linewidth]{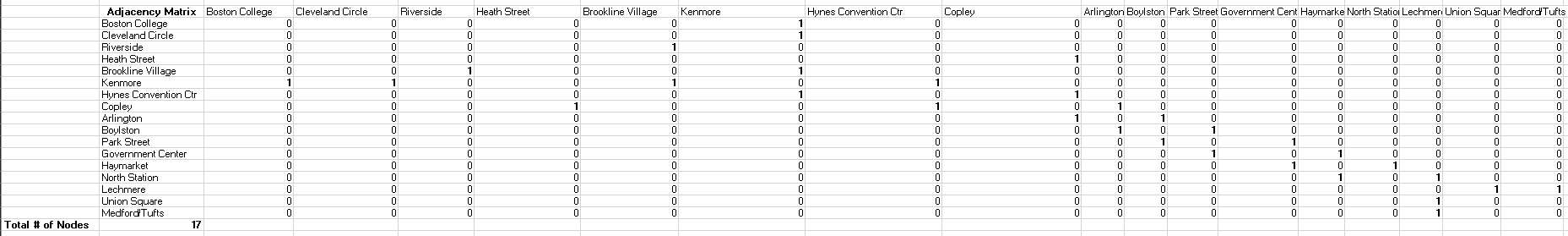}
    \caption{Adjacency Matrix}
    \label{fig:placeholder}
\end{figure}

\begin{figure}[H]
    \centering
    \includegraphics[width=1\linewidth]{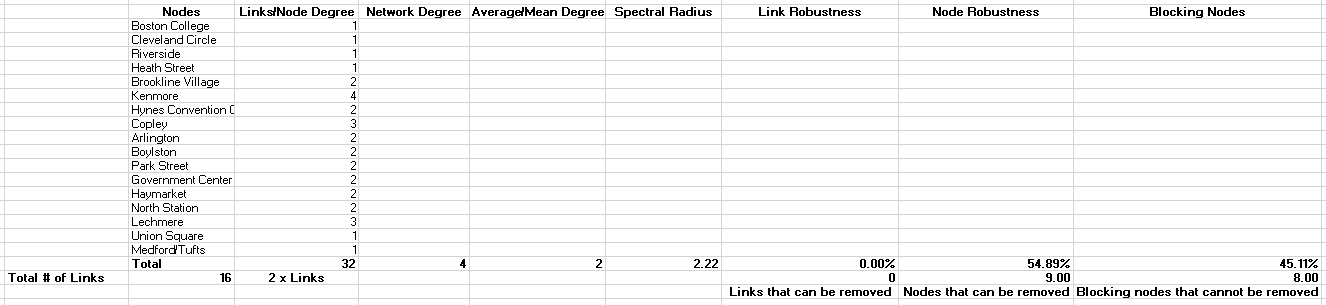}
    \caption{Metrics}
    \label{fig:placeholder}
\end{figure}

\begin{figure}[H]
    \centering
    \includegraphics[width=1\linewidth]{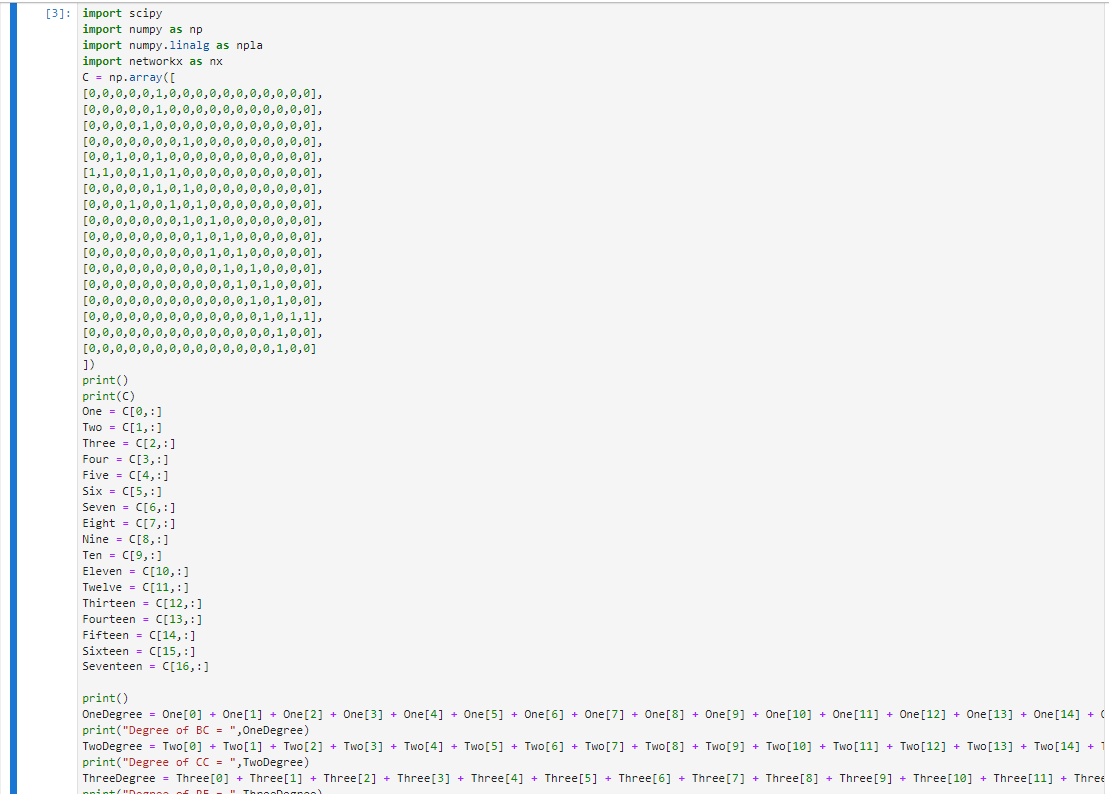}
    \caption{Python Code - Network Metrics (1/9)}
    \label{fig:placeholder}
\end{figure}

\begin{figure}[H]
    \centering
    \includegraphics[width=1\linewidth]{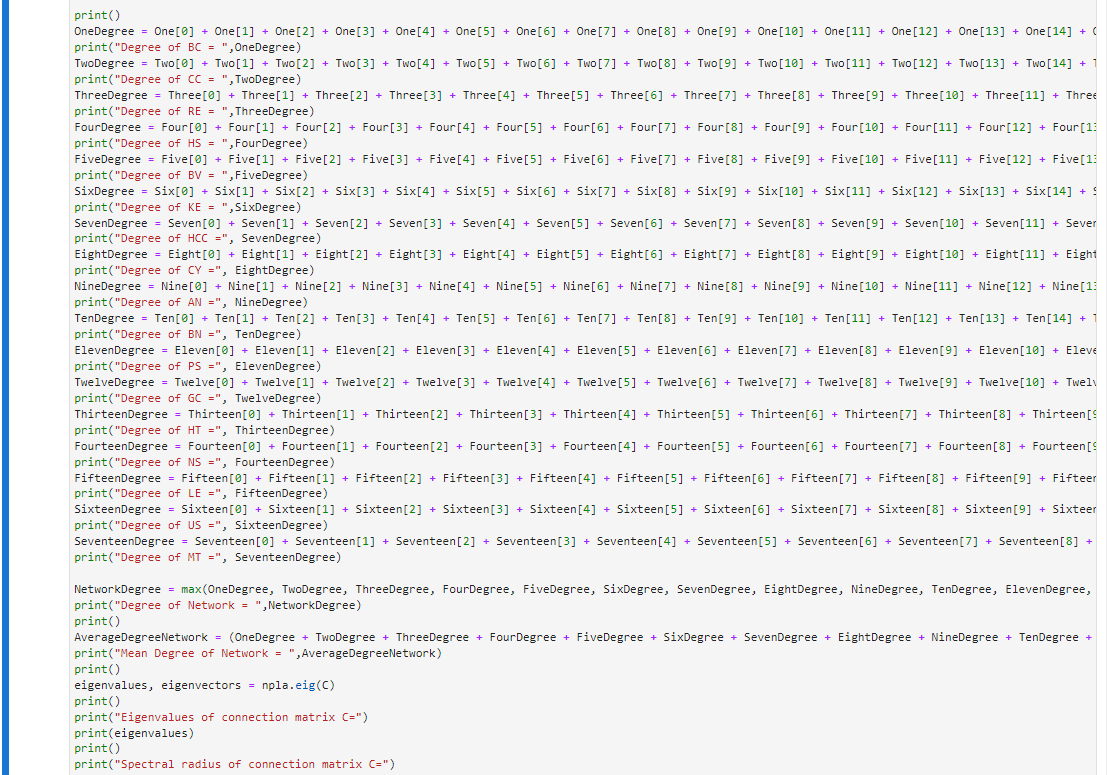}
    \caption{Python Code - Network Metrics (2/9)}
    \label{fig:placeholder}
\end{figure}

\begin{figure}[H]
    \centering
    \includegraphics[width=1\linewidth]{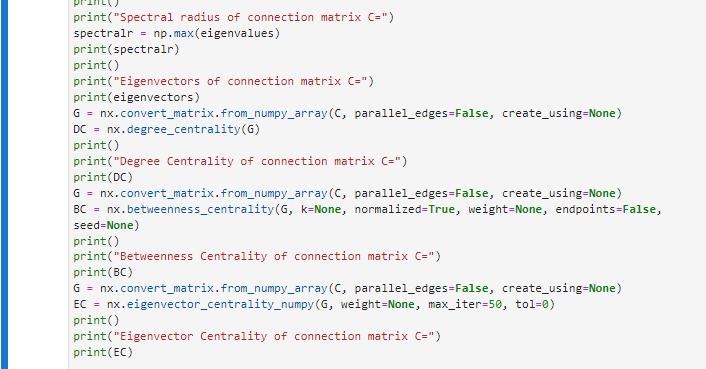}
    \caption{Python Code - Network Metrics (3/9)}
    \label{fig:placeholder}
\end{figure}

\begin{figure}[H]
    \centering
    \includegraphics[width=1\linewidth]{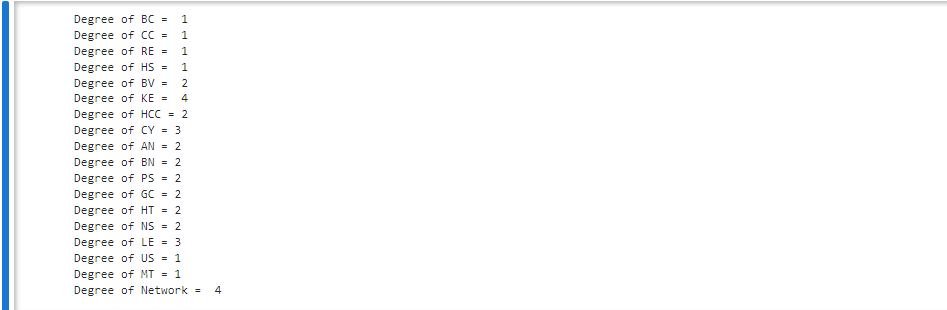}
    \caption{Python Code - Network Metrics (4/9)}
    \label{fig:placeholder}
\end{figure}

\begin{figure}[H]
    \centering
    \includegraphics[width=1\linewidth]{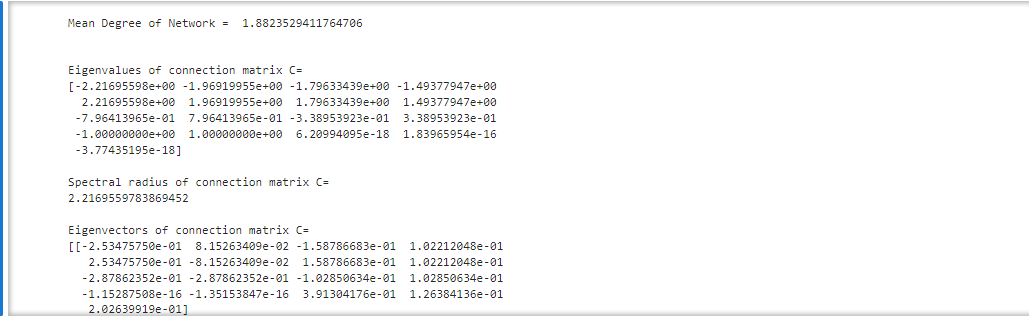}
    \caption{Python Code - Network Metrics (5/9)}
    \label{fig:placeholder}
\end{figure}

\begin{figure}[H]
    \centering
    \includegraphics[width=1\linewidth]{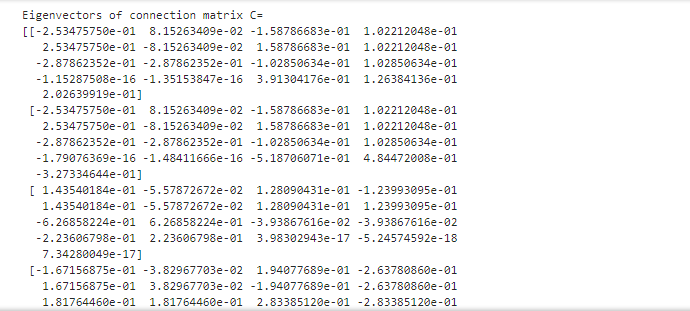}
    \caption{Python Code - Network Metrics (6/9)}
    \label{fig:placeholder}
\end{figure}

\begin{figure}[H]
    \centering
    \includegraphics[width=1\linewidth]{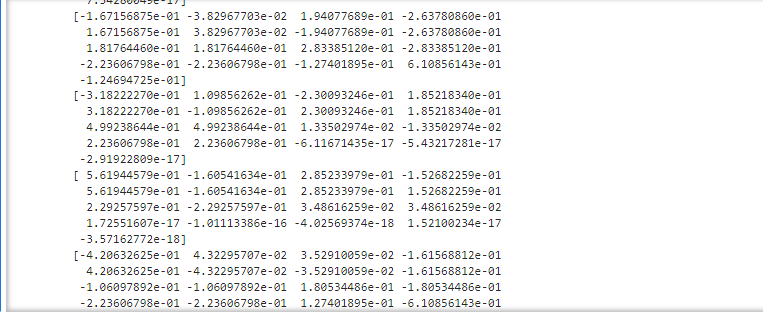}
    \caption{Python Code - Network Metrics (7/9)}
    \label{fig:placeholder}
\end{figure}

\begin{figure}[H]
    \centering
    \includegraphics[width=1\linewidth]{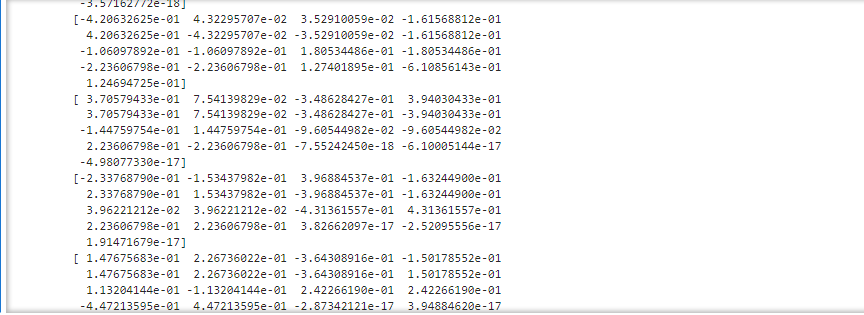}
    \caption{Python Code - Network Metrics (8/9)}
    \label{fig:placeholder}
\end{figure}

\begin{figure}[H]
    \centering
    \includegraphics[width=1\linewidth]{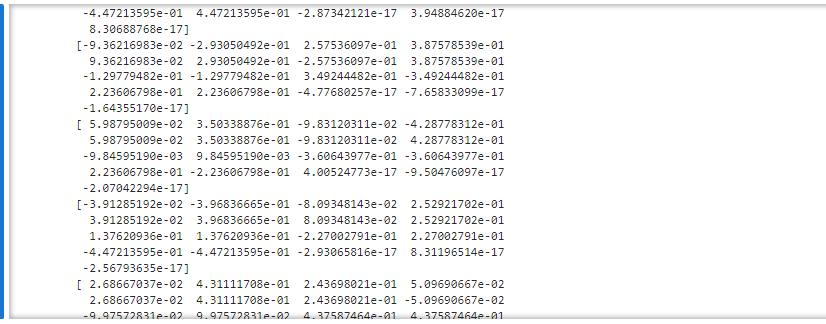}
    \caption{Python Code - Network Metrics (9/9)}
    \label{fig:placeholder}
\end{figure}

\begin{figure}[H]
    \centering
    \includegraphics[width=1\linewidth]{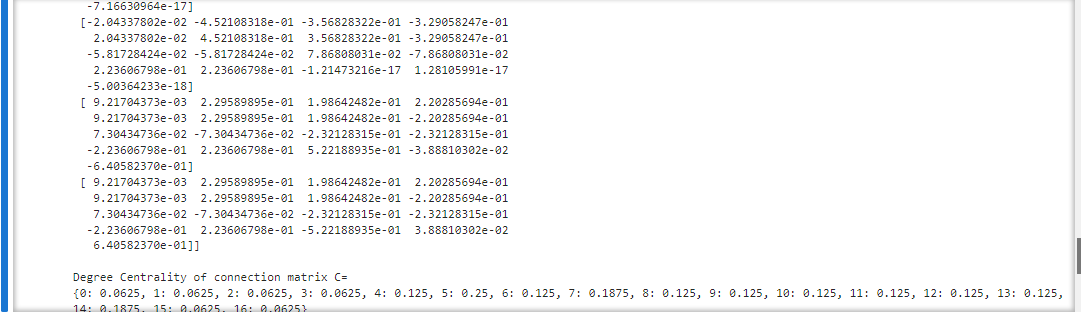}
    \caption{Python Code - Network Metrics Output (1/2)}
    \label{fig:placeholder}
\end{figure}

\begin{figure}[H]
    \centering
    \includegraphics[width=1\linewidth]{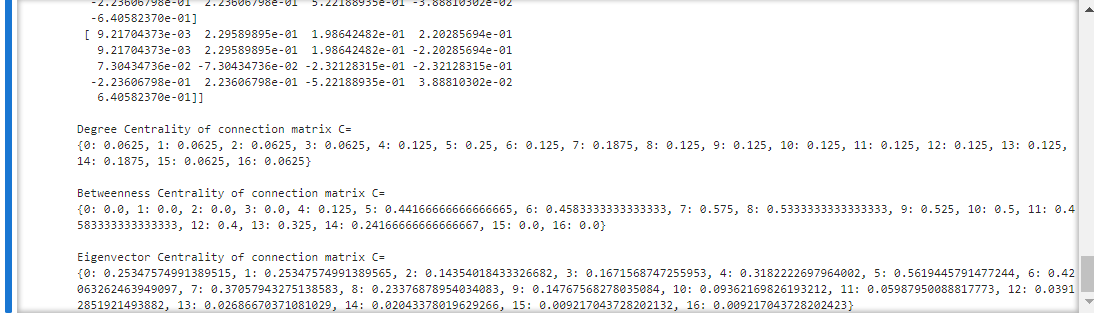}
    \caption{Python Code - Network Metrics Output (2/2)}
    \label{fig:placeholder}
\end{figure}

\newpage\textit{Calculated Risk}

\begin{figure}[H]
    \centering
    \includegraphics[width=1\linewidth]{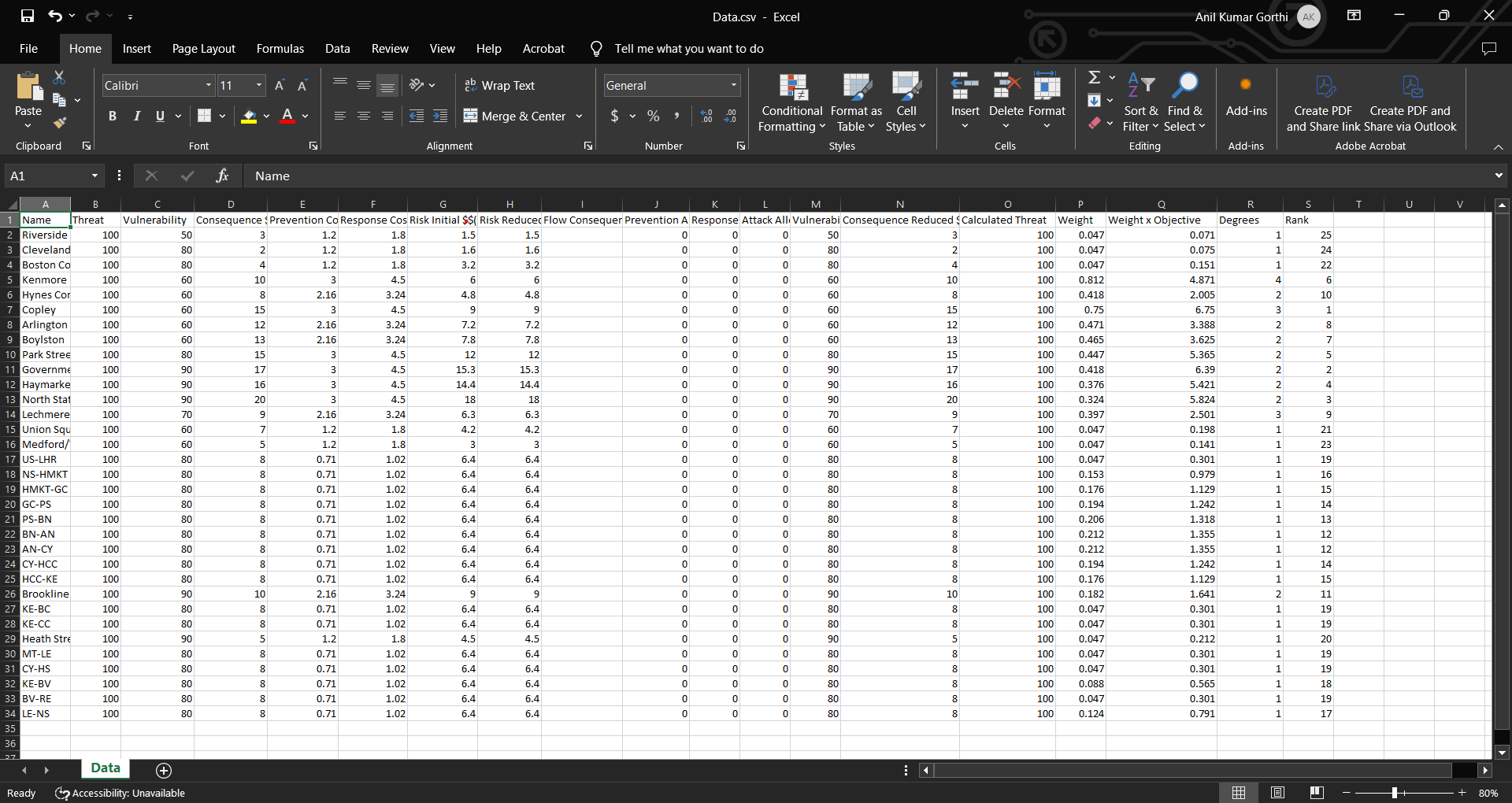}
    \caption{Network data with calculated risk (final)}
    \label{fig:placeholder}
\end{figure}

\begin{figure}[H]
    \centering
    \includegraphics[width=1\linewidth]{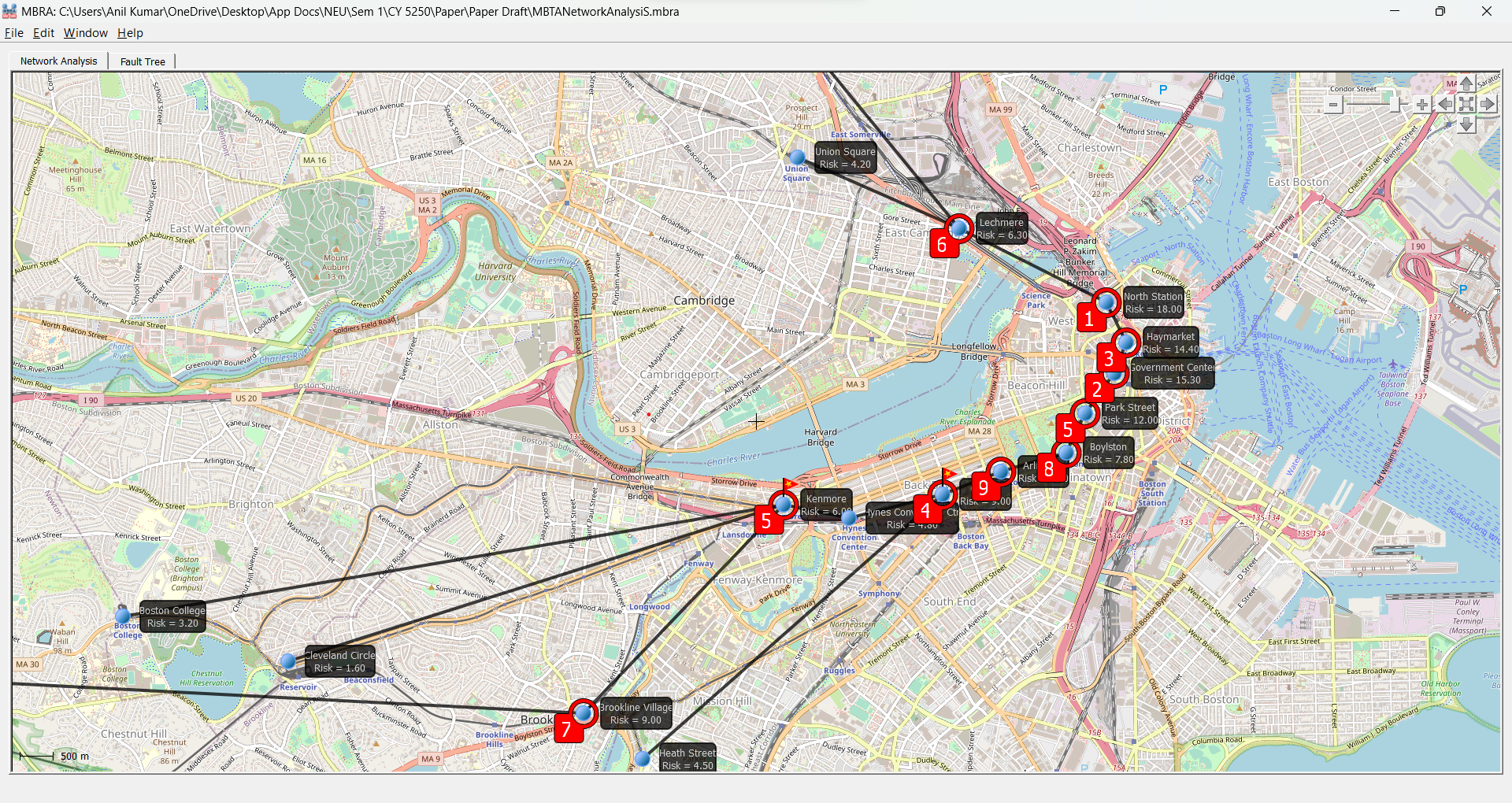}
    \caption{Network graph with calculated risk (1/2)}
    \label{fig:placeholder}
\end{figure}

\begin{figure}[H]
    \centering
    \includegraphics[width=1\linewidth]{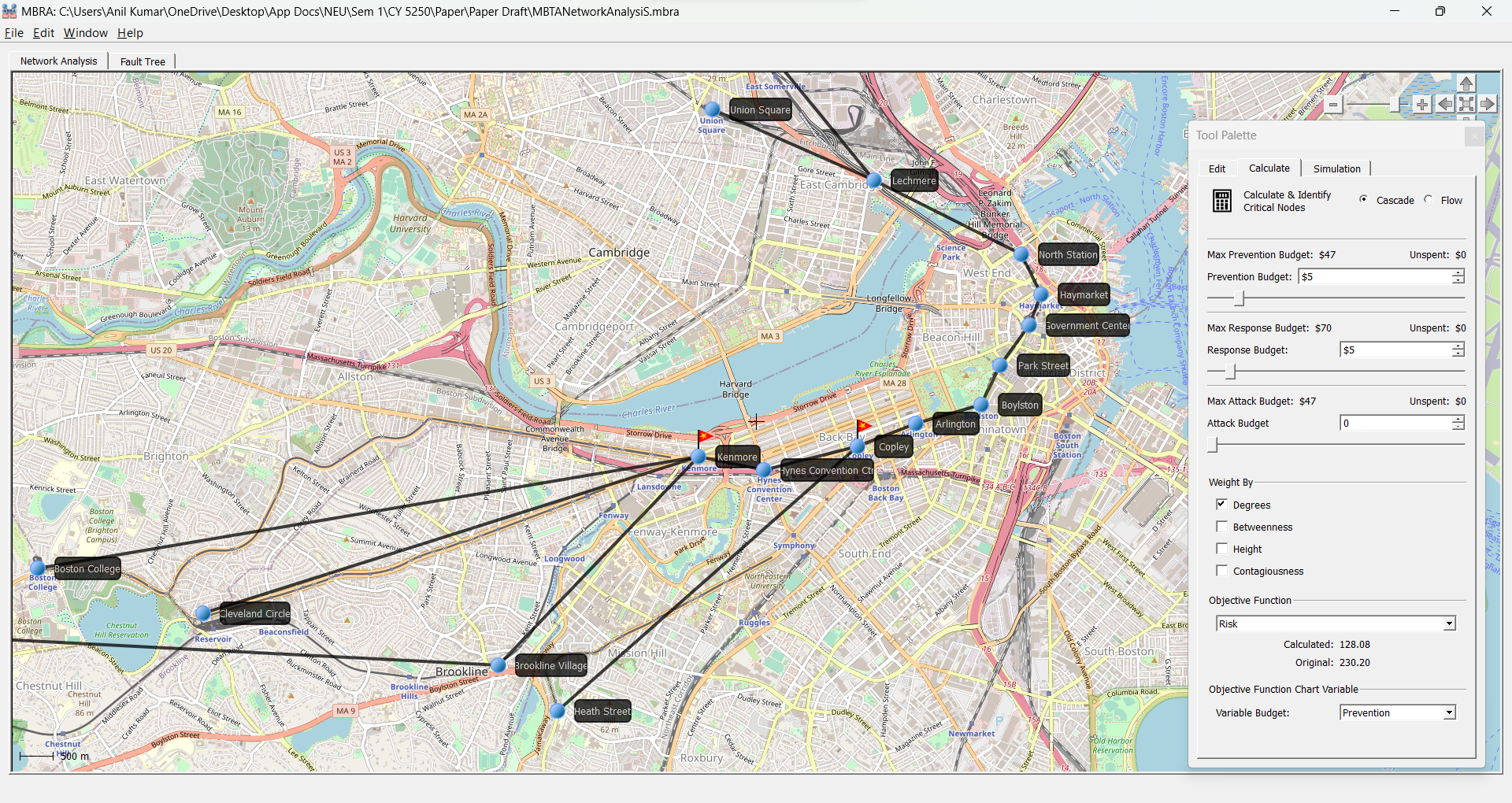}
    \caption{Network graph with calculated risk (2/2)}
    \label{fig:placeholder}
\end{figure}

\begin{figure}[H]
    \centering
    \includegraphics[width=1\linewidth]{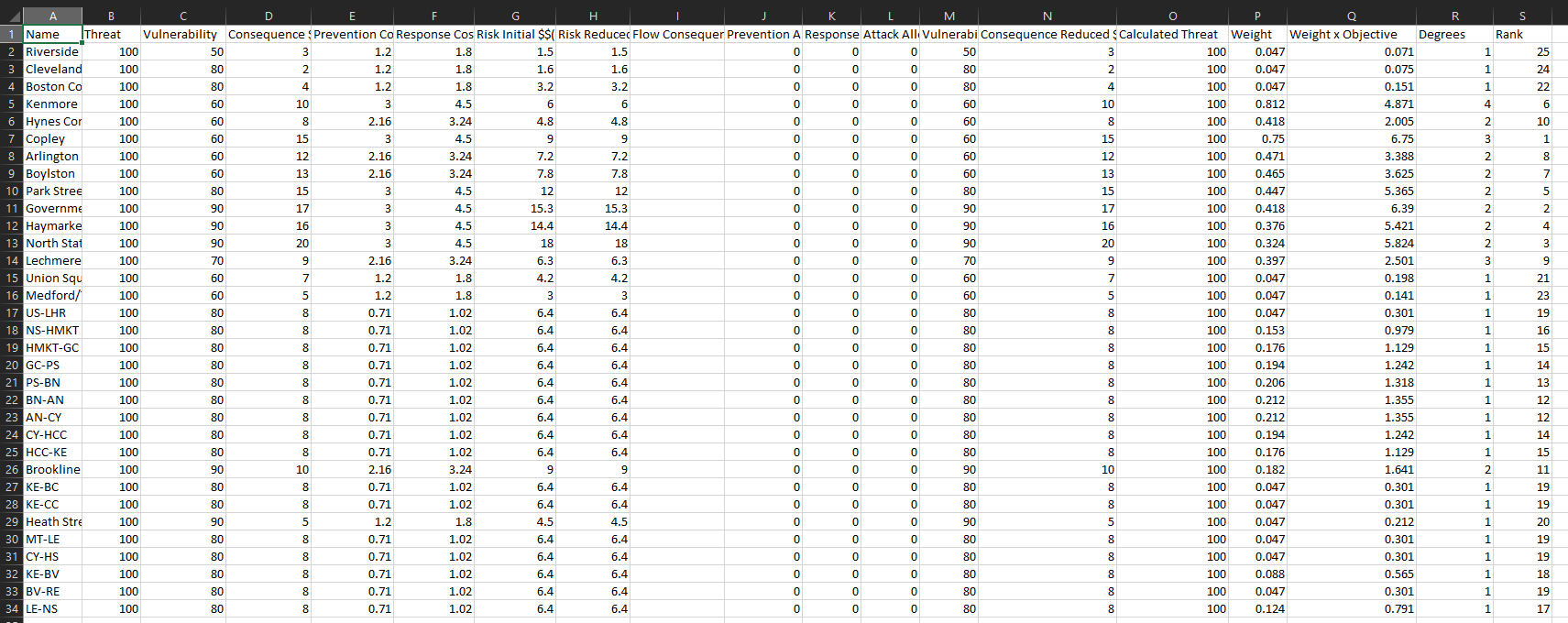}
    \caption{Network graph data table}
    \label{fig:placeholder}
\end{figure}

\begin{figure}[H]
    \centering
    \includegraphics[width=1\linewidth]{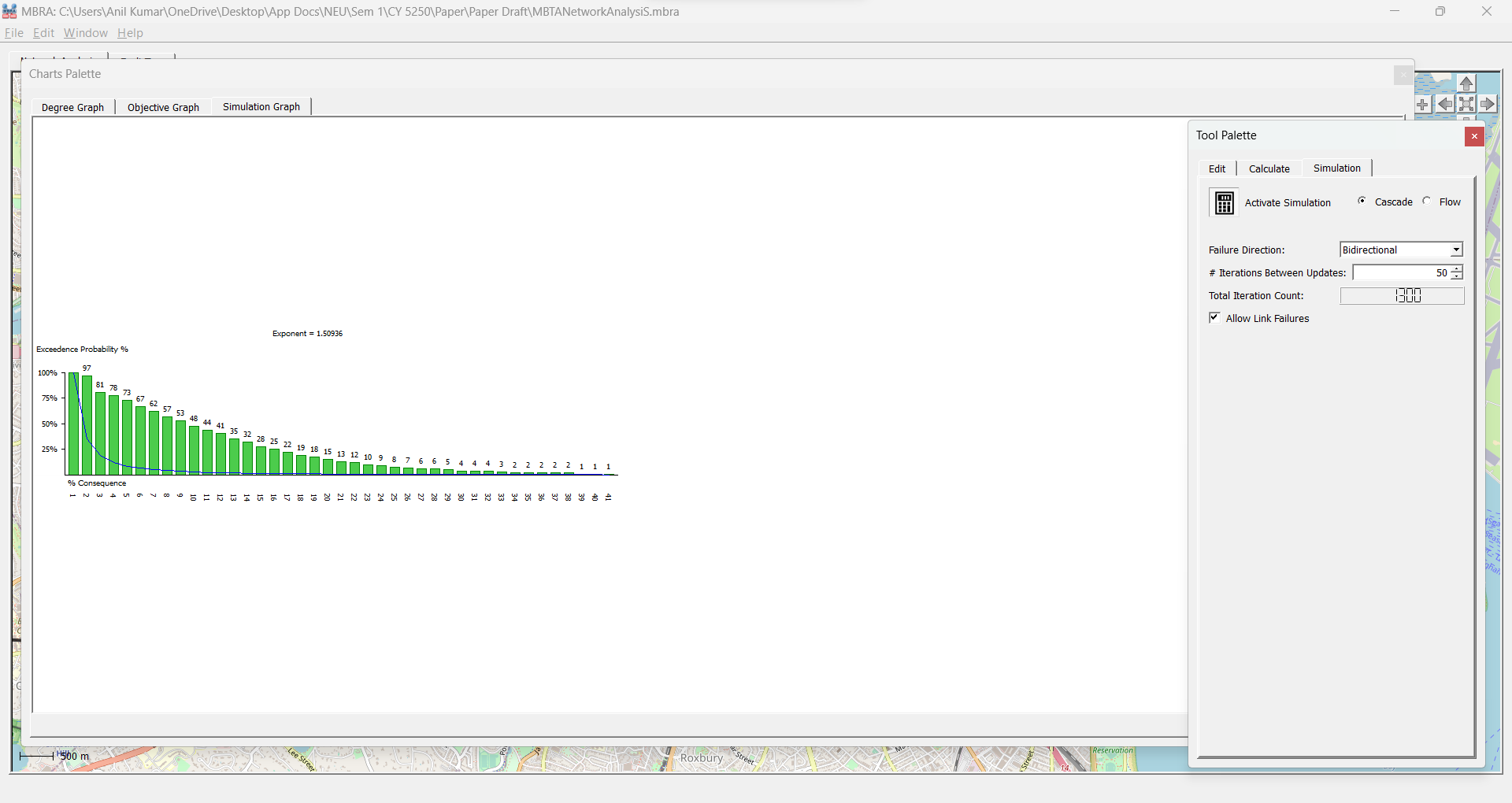}
    \caption{Simulation Graph}
    \label{fig:placeholder}
\end{figure}

\begin{figure}[H]
    \centering
    \includegraphics[width=1\linewidth]{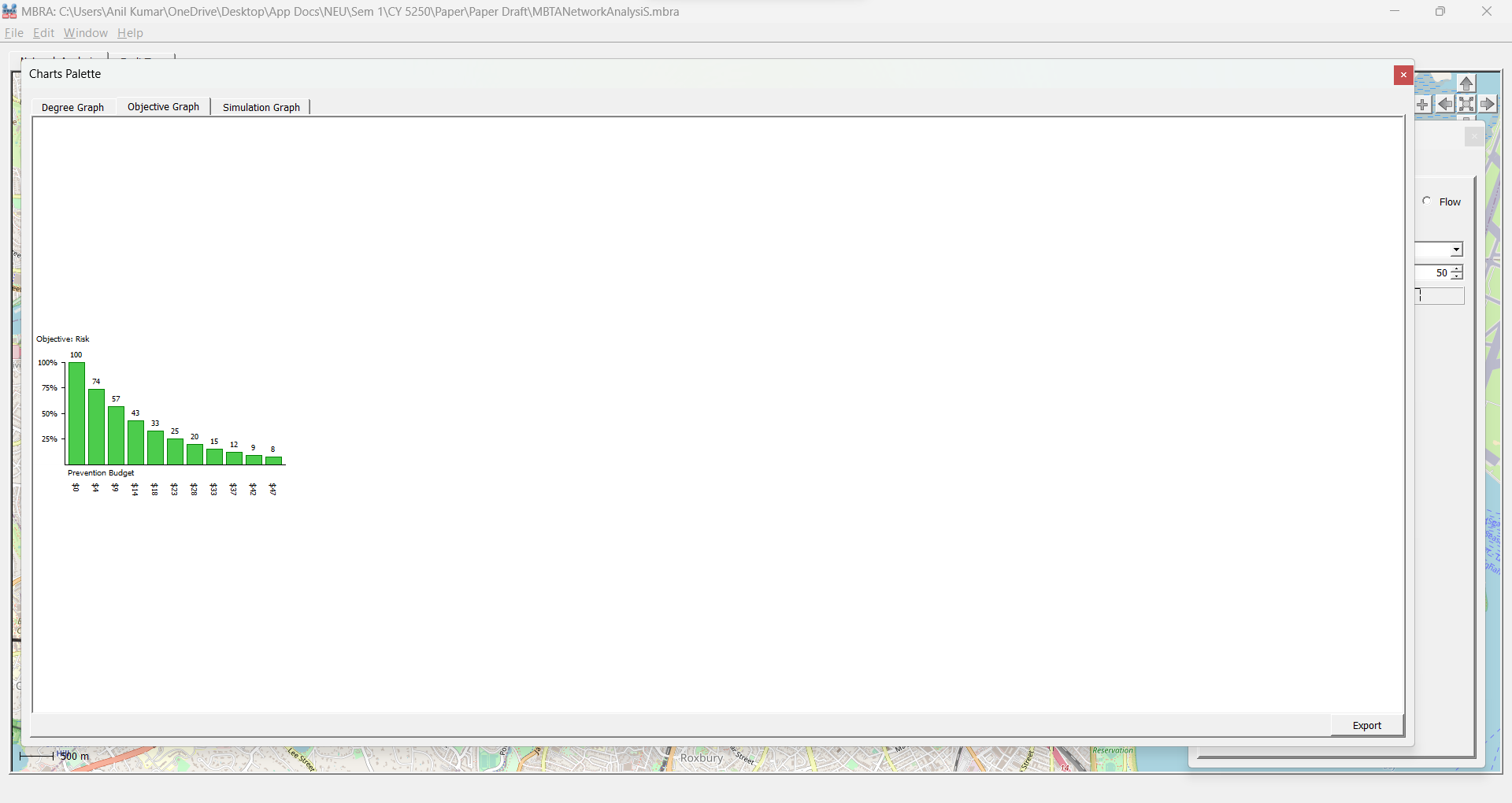}
    \caption{Objective Graph}
    \label{fig:placeholder}
\end{figure}

\begin{figure}[H]
    \centering
    \includegraphics[width=1\linewidth]{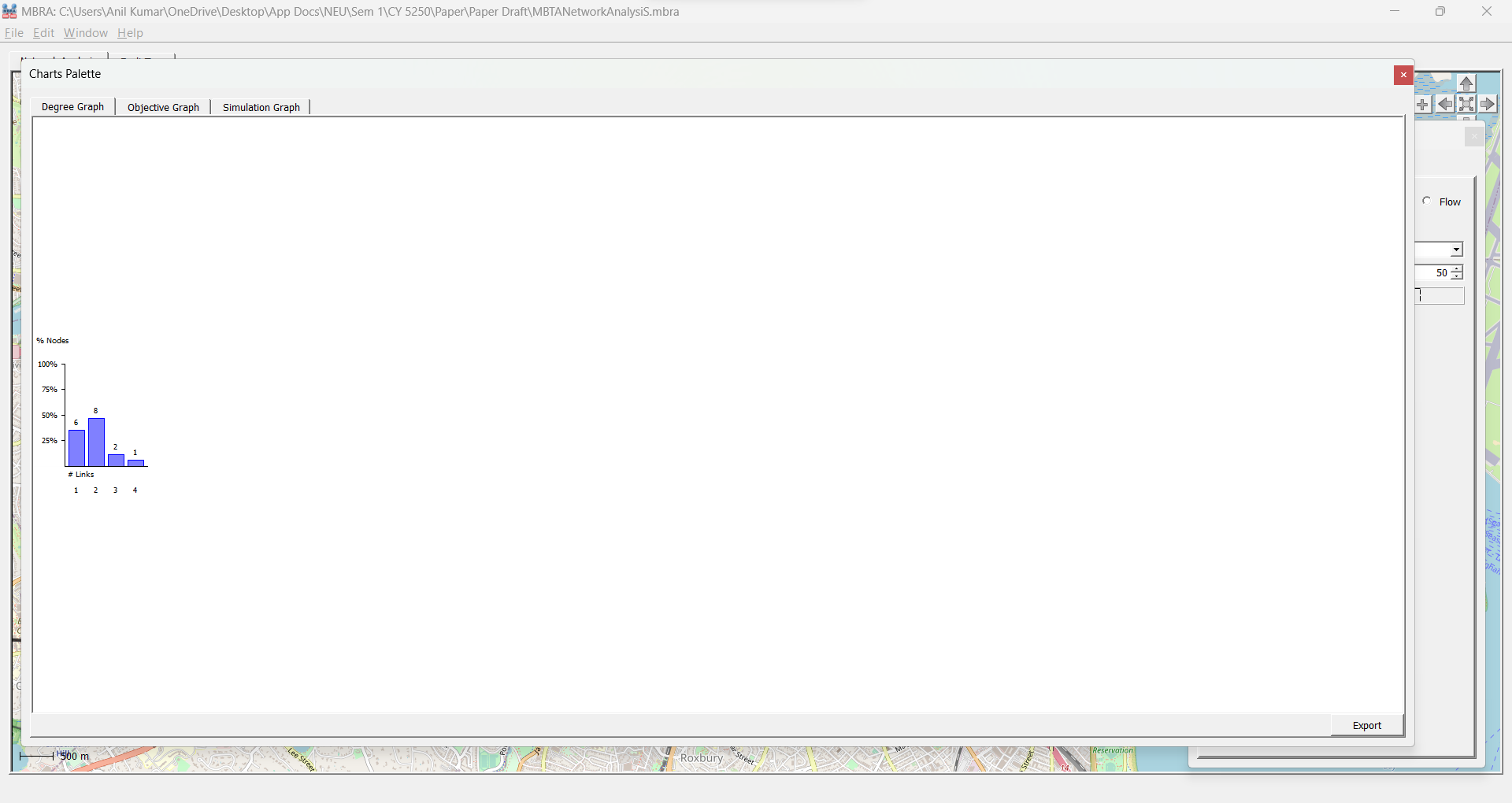}
    \caption{Degree Graph}
    \label{fig:placeholder}
\end{figure}

\begin{figure}[H]
    \centering
    \includegraphics[width=1\linewidth]{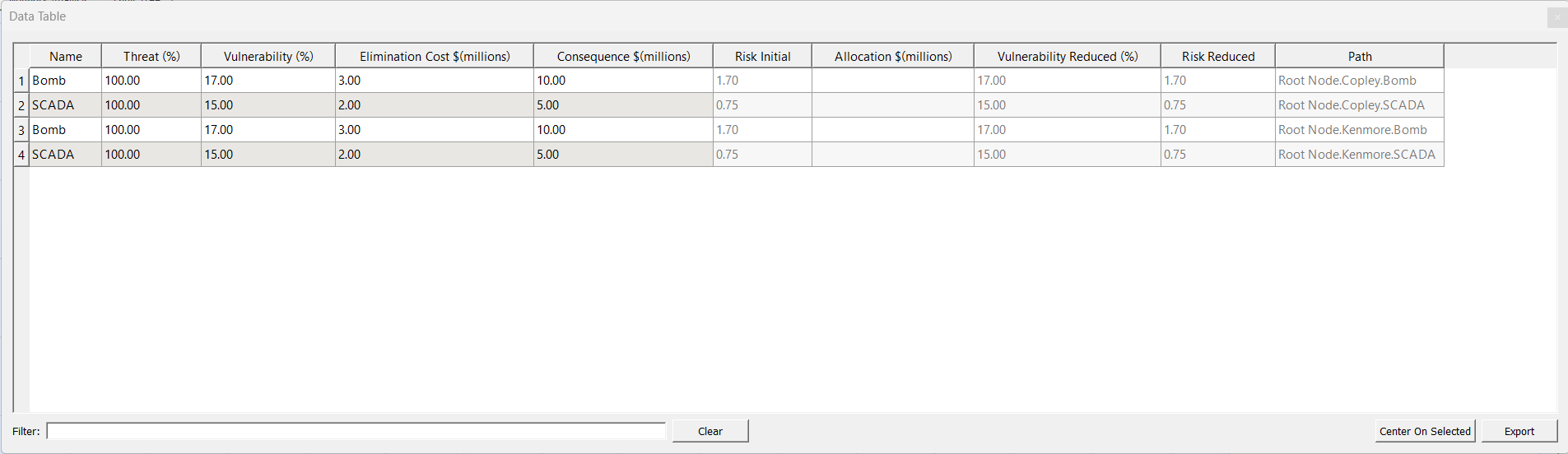}
    \caption{Fault Tree Data Table}
    \label{fig:placeholder}
\end{figure}

\end{document}